\documentclass[letterpaper,12pt]{JHEP3}
\usepackage{graphicx}
\newcommand{\beq}{\begin{equation}}
\newcommand{\eeq}{\end{equation}}
\newcommand{\barr}{\begin{eqnarray}}
\newcommand{\earr}{\end{eqnarray}}

\newcommand{\Mfunction}[1]{#1}

\def\cH{{\cal H}}

\def\cG{{\cal G}}

\def\cF{{\cal F}}

\def\cS{{\cal S}}
\def\cO{{\cal O}}
\def\cE{{\cal E}}
\def\cL{{\cal L}}

\def\pp{{\partial}}
\def\half{{\frac{1}{2}}}

\newcommand{\p}{\partial}

\def\Ord#1{{\cal O}\left( #1\right)}

\renewcommand{\Im}{{\rm Im}}

\title{Casimir Buoyancy}
\author{R.~L.~Jaffe and A.~Scardicchio \\ Center for Theoretical Physics, \\ Laboratory for
Nuclear Sciences and Physics Department\\
Massachusetts Institute of Technology \\ Cambridge, MA 02139,
USA\\ \email{jaffe@mit.edu, scardicc@mit.edu}}
\abstract{We study the Casimir force on a \textit{single} surface
immersed in an inhomogeneous medium.  Specifically we study the
vacuum fluctuations of a scalar field with a spatially varying
squared mass, $m^{2}+\lambda\Delta(x-a)+V(x)$, where  $V$ is a
smooth potential and $\Delta(x)$ is a unit-area function sharply
peaked around $x=0$.  $\Delta(x-a)$ represents a semi-penetrable
thin plate placed at $x=a$.  In the limits
$\{\Delta(x-a)\to\delta(x-a),\ \ \lambda\to\infty \}$ the scalar
field obeys a Dirichlet boundary condition, $\phi=0$, at $x=a$. We
formulate the problem in general and solve it in several
approximations and specific cases. In all the cases we have
studied we find that the Casimir force on the plate points in the
direction opposite to the force on the quanta of $\phi$: it pushes
the plate toward higher potential, hence our use of the term
\textit{buoyancy}.   We investigate Casimir buoyancy for weak,
reflectionless, or smooth $V(x)$, and for several explicitly
solvable examples.  In the semiclassical approximation, which
seems to be quite useful  and accurate, the Casimir buoyancy is a
\textit{local} function of $V(a)$.  We extend our analysis to the
analogous problem in $n$-dimensions with $n-1$ translational
symmetries, where Casimir divergences become more severe. We also
extend the analysis to non-zero temperatures.}
\preprint{MIT-CTP-3585}

\begin{document}
\section{Introduction}
\setcounter{equation}{0} \setcounter{figure}{0}

The interaction of a quantum field with a material medium
sometimes can be idealized by placing a boundary condition on the
field at the interface with the medium. Then the effects of the
medium can be interpreted as modifications of the zero point
energy of the field due to the boundary condition on the surface.
The classic example is the force between two parallel, grounded
conducting plates due to zero point fluctuations of the
electromagnetic field first discovered by Casimir\cite{casimir}.
Hence problems of this nature are known in general as Casimir
problems.  If the surface $\cS$  is embedded in an inhomogeneous
medium, then the quantum fluctuations of $\phi$ sense the
inhomogeneities and give rise to a force per unit area on the
surface even in the absence of a second surface.  We will refer to
this force as \textit{Casimir buoyancy}.  As ``buoyancy'' implies,
the force is opposite to the force that acts on the quanta of the
fluctuating field, at least in the cases we have been able to
study.

In this paper we examine the problem of Casimir buoyancy.  We
formulate and study the problem for the simplest possible cases.
We consider a scalar  quantum field, $\phi$, obeying the boundary
conditions imposed by a sharply peaked background on an $n-1$
dimensional hyperplane embedded in $n$-dimensional Euclidean
space.  We assume that the mass of the scalar field varies in the
direction normal to the surface, \textit{i.e.} we introduce an
interaction $\half V(x_{\perp})\phi^{2}$ into the Lagrangian,
\begin{equation}
\cL = \half\p_{\mu}\phi \p^{\mu}\phi-\half m^{2}\phi^{2}-\half
V(x_{\perp})\phi^{2}-\frac{1}{2}\lambda\Delta(x_{\perp}-a)\phi^2
\label{eq0.1}
\end{equation}
The function $\Delta$, normalized to $\int dx\Delta(x)=1$, is
assumed to be sharply peaked at $x_{\perp}=a$.  Usually a Dirac
$\delta$-function will do. We are particularly interested in the
``Dirichlet limit'', where the coupling $\lambda$ goes to infinity
and the field obeys the Dirichlet condition, $\phi=0$, on the
hyperplane.  It is easy to imagine problems to which such a
formulation applies, where a quantum field is subject to forces on
two different scales, forces at a high energy scale that can be
idealized as a boundary condition, and forces of order the mass of
$\phi$ that can be regarded as a smoothly varying
background.\footnote{In the \emph{real} Casimir effect the
idealization of the influence of the metal or dielectric on the
electromagnetic field as a static background must be taken with
some care. In particular some divergencies arising from this
idealization could be absent if a dynamic description of the
material is adopted \cite{Jaekel}.} We believe that similar
considerations apply to a gauge vector (\textit{e.g.}
electromagnetic) field in an inhomogeneous medium and to fermion
fields.  The generic problem of a quantum field constrained on a
surface and modulated by other forces in the bulk also arises in
brane world scenarios, where similar effects should also be
expected. We have not considered buoyancy in a curved space-time
where the effect could arise from an inhomogeneous curvature
\cite{BirrelDavis}.

As usual in Casimir physics,  there is little intuition to guide
us \textit{a priori}.  For example, there is no reason to expect
the buoyancy force at a point $x$ to depend only on $V$ at the
point $x$.  In general we find that the buoyancy depends
non-locally on $V$.  However when the background field is smooth
enough to admit a WKB approximation we find that the buoyancy
reduces to a local function of the background field. Likewise we
know of no argument to give the sign of the Casimir buoyancy.
Should it be parallel to the force on the quanta of $\phi$, or
antiparallel?  In all the examples we have been able to study we
find that the force is opposite to the force on the quanta, hence
a buoyancy.

Casimir problems can suffer from divergences of two different
kinds\cite{dirichlet}.  The first are the familiar divergences
that afflict any quantum field theory.  The Casimir energy for a
fluctuating field $\phi$ in a time-independent
background\footnote{For us $\sigma(x)=\lambda\delta(x-a)+V(x)$.}
$\sigma(x)$ is the full one-loop effective energy, $E[\sigma]$,
the sum over all one loop Feynman diagrams with arbitrary
insertions of $\sigma$, as shown in the first line of
Fig.~\ref{effectiveenergy}\cite{eft}.  The low order Feynman
diagrams diverge:  the 1-point function diverges for $n\ge 1$, the
two point function for $n\ge 3$, \textit{etc.} As usual, these
divergences are cancelled by counterterms that are polynomials in
$\sigma$, $\cL_{\rm CT}=c_{1}\sigma +\half c_{2}\sigma^{2}+\ldots$
(some are shown in Fig.~\ref{effectiveenergy}).  The Casimir
energy becomes dependent on the renormalized parameters of the
$\sigma$-field dynamics. For example the need for the counterterm
$\half c_{2}\sigma^{2}$ in three dimensions generates a dependence
on the renormalized $\sigma$-mass.

The second type of divergence is more interesting and more
challenging.  A realistic material medium cannot be idealized by a
boundary condition at all energy scales.  When the frequency of
the fluctuating field is high compared to the natural scale of the
interactions that characterize the material, its effects fade
away.  A background that constrains \textit{all} modes of a
fluctuating field  is unphysical, and can introduce divergences
into physical observables that cannot be removed by standard
renormalization methods\cite{Graham:2003ib}.  The origin of these
divergences is quite clear from the Feynman diagrams.  Each
diagram involves integrals over the momenta carried by the
external lines, $\sigma(x)=\int dpe^{-ipx}\hat \sigma(p)$.  If the
background has a discontinuity in some derivative, say the $k$-th,
then at large $p$, $\hat \sigma(p)\sim p^{-k}$ and the integrals
diverge for sufficiently large number of dimensions. There are no
renormalization counter terms available to cancel these
divergences.  They signify that the quantity under consideration,
even a directly measurable one like the Casimir force, is
sensitive to the high energy cutoffs, $\Omega$, above which the
material no longer affects the field (say $\hat\sigma(p)=0$ for
$p>\Omega$).

In a given model there may be some quantities which admit a
Casimir (\textit{i.e.}\ boundary condition) description and others
that do not. The boundary condition idealization shares with the
effective field theory the notion of a separation of scales. The
material structure is characterized by a high energy scale, the
cutoff $\Omega$.  Modes with energies below this scale obey a
boundary condition, modes with energies at or above $\Omega$ do
not.  If the boundary condition method is applicable, then physics
at energies much lower than $\Omega$ can be described by the
boundary condition without reference to $\Omega$ at all.  For
Casimir's original problem, parallel conducting plates, the plasma
frequency $\omega_{\rm plasma}$ sets the high energy scale
$\Omega\sim \hbar\omega_{\rm plasma}$, and the plate separation,
$d$, or rather $\hbar c/d$, is the energy scale of physical
interest. When $\omega_{\rm plasma}\gg c/d$ the force between
parallel plates is well described by the boundary condition
calculation. It can be shown that the Casimir idealization works
for Casimir forces between rigid bodies in vacuum and for local
observables like the energy density outside the material (at
distances greater than $\hbar c/\Omega$), two examples of immense
practical importance. Other observables are not so fortunate.  It
was shown in Ref.~\cite{Graham:2003ib} that the Casimir
\textit{pressure} on a sphere of radius $R$, for example, depends
on the cutoff, so it is not possible to study the pressure, even
when $\hbar c/R \ll \Omega$, without characterizing the material
in detail.

It is important to learn the circumstances under which there is an
effective low energy description of buoyancy independent of the
material cutoffs.  We find that the answer to this question
depends on the number of dimensions, $n$.  In one dimension, where
the surface is a point, there are no divergences of any kind and
the Casimir buoyancy is independent of the cutoffs.  Thus, for
example, it goes to a finite limit as $\lambda\to\infty$.  Higher
dimensions can be studied relatively easily using the formalism
developed in Refs.~\cite{Graham:2001dy}.  For $n<2$ the situation
is the same as for $n=1$.  For $2\le n<3$, the Casimir buoyancy
remains independent of the details of the surface, but the limit
$\lambda\to\infty$ cannot be taken (it diverges like
$\lambda^{n-2}$).  For $n\ge 3$ there is no separation of scales.
The buoyancy depends on the details of the structure of the
material.

\FIGURE{
\includegraphics[width=12cm]{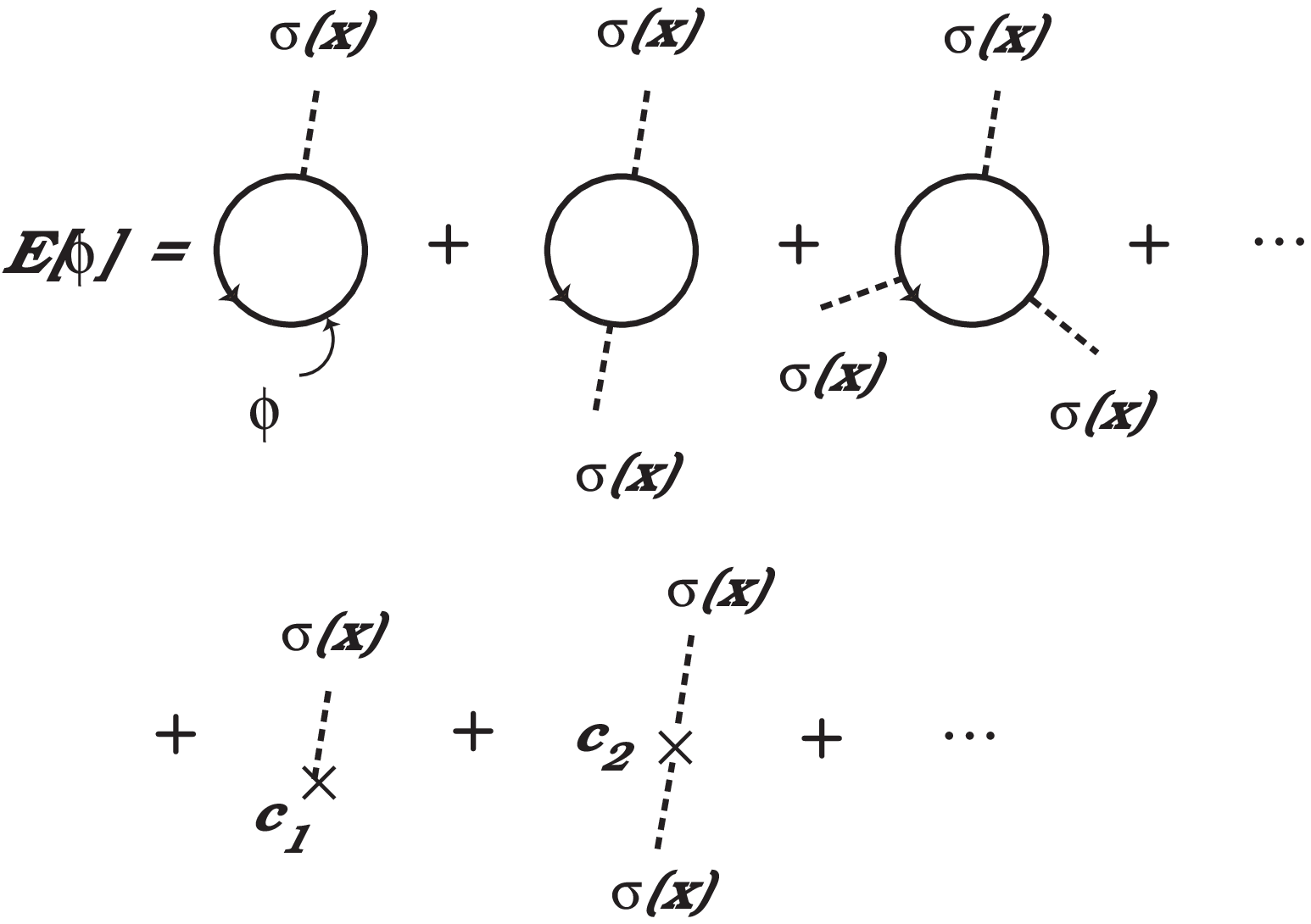}
\caption{\label{effectiveenergy}\sl The Casimir energy of a
fluctuating field, $\phi$, coupled to a time independent
background field, $\sigma$, via ${\cal
L}_{I}=\half\sigma(x)\phi^{2}$ is proportional to the sum of all
one loop diagrams.  The sum must include the contributions of
counter terms, polynomials in $\sigma$, required to cancel the
loop divergences.  The structure of the counter terms depends on
the number of dimensions, $n$.  The second line in the figure
shows the counter terms required in three dimensions where the 1-
and 2- point functions are primitively divergent. } }

In the next section we describe the formulation of the Casimir
buoyancy problem in one dimension.  We recast it in terms of the
Schr\"odinger equation Greens function with potential $V(x)$, and
express the buoyancy in terms of the bound and scattering states
of $V$.  We study both fixed $\lambda$ and the limit
$\lambda\to\infty$.  In Section III we describe some important
approximations and special cases:  For deep and smooth $V(x)$ we
derive a WKB approximation; we study the case where $V(x)+m^2=0$
(a turning point) which cannot be analyzed by means of WKB,
proving buoyancy in this case as well. We study the Casimir
buoyancy force in a thermal state finding the buoyancy is
 qualitatively  not affected by a non-zero temperature.
 For a reflectionless
potential we show that only the bound states matter. We study the
buoyancy outside the range of $V(x)$ and construct the first Born
approximation as an example of non-local but simple result. In
Section IV we go through some explicit, solvable examples. Then in
Section V we generalize from $n=1$ to higher dimensions.

\section{Formulation of the problem}

The situation of interest is summarized by eq.~(\ref{eq0.1}).  It
might arise if a quantum field is coupled to one field
characterized by a high mass scale that lives on the hyperplane
$x_{\perp}=0$, and to another field $V$ characterized by a lower
energy scale.  As usual we take $V$ to be time independent and
externally determined; we ignore the back reaction of $\phi$ on
$V$.  The principal dynamical effects occur in the one non-trivial
direction along which $V$ varies, so we begin by studying the
one-dimensional problem.

\subsection{General considerations}

In one dimension the Lagrangian for $\phi$ reads
\begin{equation}
    \cL = \half \dot\phi^2-\half\phi'^2 -\half\left( m^2+\lambda\Delta(z-a)+
    V(z)\right)\phi^2
    \label{eq1}
\end{equation}
We are interested in the vacuum energy of the $\phi$-field as a
functional of $\Delta$ and $V$,
\begin{equation}
    \cE=\half\hbar\ \int\!\!\!\!\!\!\!\!\sum \omega =\frac{1}{2}\hbar\int_0^\infty dE \sqrt{E}\frac{dN}{dE},
    \label{eq2}
\end{equation}
a sum over the discrete spectrum, if any, and integral over the
continuum, where the $\{\omega^{2}\}$ are the eigenfrequencies of
the Schr\"odinger Hamiltonian,
\begin{equation}
    \cH = -d^{2}/dx^{2} +  m^2 +\lambda\Delta(x-a)+V(x),
    \label{eq3}
\end{equation}
which will,  in general,  range over both discrete and continues
values.\footnote{We set $\hbar=c=1$ until further notice.} So the
problem can be recast in the form of a Schr\"odinger equation,
$\cH\phi=E\phi$, with $E=\omega^{2}\equiv k^{2}+m^{2}$. If we
define $\cG(x,x',E)=\langle x'|\frac{1}{\cH-E}|x\rangle$ to be the
equivalent Schr\"odinger Greens function, then we can write the
density of  states, $dN/dE$, as
\begin{equation}
    \frac{dN}{dE}=\frac{1}{\pi}\ {\rm Im}\int d^{\ \!n}x\,\cG(x,x,E+i\epsilon)
    \label{eq5}
\end{equation}
where $n$ is the number of spatial dimensions. The $i\epsilon$
displaces the poles in $\cG$ into the lower half $E$-plane, and
insures that the propagator, $\cG(x,x',t)=\int dE
e^{-iEt}\cG(x,x',E+ i\epsilon)$, is causal. Substituting for the
density of states into eq.~(\ref{eq2}), the Casimir energy can be
written
\begin{equation}
    \cE=\frac{1}{2\pi}{\rm Im}\int_{0}^{\Omega} dE\sqrt{E}\int d^{\ \!n}x  \cG(x,x,E+i\epsilon),
    \label{eq5.1}
\end{equation}
where we have assumed that the spectrum is positive
definite\footnote{Negative energy single particle states in an
external field correspond to vacuum instabilities that we do not
consider here.} and we have introduced a cutoff $\Omega$ above
which the boundary must be characterized in greater detail. We
already discussed in the Introduction the conditions (on $n$ and
$\sigma=\lambda\Delta+V$) under which we can take
$\Omega\to\infty$ without affecting the low energy physics. We
will go over some these arguments again in Section
\ref{sec:ngtr1}. At the moment we assume all these conditions to
be fulfilled so we can take the limit $\Omega\to\infty$ without
placing further restrictions on $\sigma(x)$.

\subsection{Force on a sharp surface}

We are interested in the case where $\Delta(x-a)$ is  a Dirac
$\delta$-function, $\Delta(x)= \delta(x)$, and $\lambda>0$ since
this repels the field $\phi$ from the ``surface'' $x=a$.  If we
restrict the analysis to $1$ spatial dimension (the extension to
$n>1$ with translational symmetry in $n-1$ dimensions will be
performed in Section \ref{sec:ngtr1}) it is possible to express
the Greens function $\cG$ in terms of the (simpler) Greens
function, $\cG_{0}(x',x,E) \equiv\langle
x'|\frac{1}{\cH_{0}-E}|x\rangle$ in the presence of $V(x)$ alone,
\begin{equation}
    \cH_{0}=  -\frac{d^{2}}{dx^{2}}+m^{2}+V(x).
        \label{eq7}
\end{equation}
First, we write the Lippmann-Schwinger equation\footnote{Here we
are considering a fully renormalized Hamiltonian $\cH$, a function
of renormalized masses and couplings.} for $\cG$ in terms of
$\cG_{0}$ and $\Delta(x)$,
\begin{equation}
    \cG(x,x',E)=\cG_{0}(x,x',E)-\lambda\int dy \cG_{0}(x,y,E)\Delta(y-a)
    \cG(y,x',E)
    \label{eq8}
\end{equation}
When $\Delta(x-a)\to\delta(x-a)$ it is easy to solve for
$\cG$,\footnote{After this paper was completed Brian Winn, in a
conversation with one of us (AS), pointed out that singularly
perturbed Hamiltonians like (\ref{eq3}) have been studied in
chaotic billiards theory where they are called \emph{\v Seba
billiards} and generalizations of Eq.\ (\ref{eq9}) can be found in
the literature on this subject \cite{Zorbas}.}
\begin{equation}
    \cG(x,x',E)=\cG_{0}(x,x',E)-\lambda\frac{\cG_{0}(x,a,E)
    \cG_{0}(a,x',E)}{1+\lambda \cG_{0}(a,a,E)}
    \label{eq9}
\end{equation}
For the density of states, we require the integral over $x$ of
$\cG(x,x,E)$. Using the identity
\begin{eqnarray}
    \int dx\cG_{0}(x,a,E)\cG_{0}(a,x,E)&=&
    \int dx \langle x|\frac{1}{\cH_{0}-E}|a\rangle
    \ \langle a|\frac{1}{\cH_{0}-E}|x\rangle\nonumber\\
    &=& \langle a|\frac{1}{(\cH_{0}-E)^{2}}|a\rangle ,\nonumber\\
    &=&\frac{\pp}{\pp E}\cG_{0}(a,a,E)
    \label{eq10}
\end{eqnarray}
we obtain
\begin{eqnarray}
    \int dx \cG(x,x,E)&=&\int dx  \cG_{0}(x,x,E)
    -\frac{\lambda}{1+\lambda\cG_{0}(a,a,E)}\frac{\pp}{ \pp E}
    \cG_{0}(a,a,E) \nonumber\\
    &=&\int dx  \cG_{0}(x,x,E)
    -\frac{\pp}{\pp E}\ln\left(1+\lambda\cG_{0}(a,a,E)\right)
        \label{eq11}
\end{eqnarray}
Substituting into eq.~(\ref{eq5.1}), we find
\begin{equation}
    \cE = -\frac{1}{2\pi}{\rm Im}\int_{0}^{\Omega} dE\sqrt{E}
    \frac{\pp}{\pp E}\ln\left(1+\lambda\cG_{0}(a,a,E+i\epsilon)\right)
    \label{eq12}
\end{equation}
where we have dropped a term  of the form
$\cE_{0}=\frac{1}{2\pi}{\rm Im}\int_{0}^{\Omega}dE\sqrt{E}\int
dx\cG_{0}(x,x,E+i\epsilon)$, which does not  depend on $a$ and
therefore does not  contribute to  the Casimir force, $-d\cE/da$.
Since we are only interested in the force, not the energy, we
differentiate with respect to $a$, $\cF=-\pp \cE/\pp a$,
\begin{equation}
    \cF =  \frac{1}{2\pi}{\rm Im}\int_{0}^{\infty} dE\sqrt{E}
    \frac{\pp^{2}}{\pp E\pp a}\ln\left(1+\lambda\cG_{0}(a,a,E+i\epsilon)\right)
    \label{eq12.1}
\end{equation}
where we have taken $\Omega\to\infty$ since according to our
previous discussions the limit exists and is finite.

The analytic structure of the integrand of eq.~(\ref{eq12.1}) is
important for our analysis (see Figure \ref{complexplane}) because
only the imaginary part contributes to $\cF$, and the integrand is
real for real $E$ except at its singularities.   The $\sqrt{E}$
gives a branch cut running from $E=0$ to $\infty$, which we place
along the negative real axis. For real, positive $E$ there are two
regions of interest.  Above threshold for scattering, $E>m^{2}$,
$\cG_{0}$ is complex, so the integrand has a cut with branch point
at $E=m^{2}$.  For $E$ real and below threshold $\cG_{0}(a,a,E)$
is real, so the only singularities occur when the argument of the
logarithm, $1+\lambda \cG_{0}(a,a,E+i\epsilon)$, vanishes, where
the integrand has poles.  When $E$ is below the spectrum of
$\cH_{0}$, $\cG_{0}(a,a,E)$ is positive and
$1+\lambda\cG_0(a,a,E)$ cannot vanish.  So there are no
singularities in the domain $0<E<m^{2}$ unless $\cH_{0}$ has bound
states.  Suppose, then, that $\cH_{0}$ has bound states at
$E_{1}$, $E_{2}$, $E_{3}$, ... $E_{M}$. Because $\cG_{0}\sim
1/(E_{j}-E)$ near the $j^{\rm th}$ bound state, is easy to see
that $1+\lambda\cG_{0}(a,a,E)$ must vanish at a value of E between
the $j^{\rm th}$ and $j+1^{\rm th}$ bound states of $\cH_{0}$.  At
each of these energies there is a contribution to the imaginary
part from the $i\epsilon$ in the argument of $\cG_{0}$.  Therefore
at least $M-1$ poles contribute to $\cF$.  If the pole just above
$E_{M}$ occurs at an energy below $m^{2}$ then there is one more.
These contributions have a simple physical interpretation:  They
are the contributions to the Casimir force from the bound states
in $V(x)$ subject to the boundary condition $\Delta
\psi'(a)=\lambda\psi(a)$. The analytic structure of the integrand
of eq.\ (\ref{eq12.1}) in the complex $E$-plane is summarized in
Fig.~(\ref{complexplane}).

From these considerations it is clear that the problem is
simplified if we rotate the integration contour to the negative
imaginary axis.  Making the obvious analogy to Feynman diagram
methods, we refer to this as ``Wick rotation'' to the ``Euclidean
form'' of the Casimir buoyancy.  There is no contribution to the
force from the semi-circle at large $|E|$, because for $E\gg V$ we
have $\cG_{0}(a,a,E)\sim \frac{i}{2\sqrt{E}}$. Although this
yields a contribution to $\cE$ (logarithmically divergent in the
cutoff $\Omega$), it is independent of $a$, and therefore does not
affect the force.
\FIGURE{
\includegraphics[width=12cm]{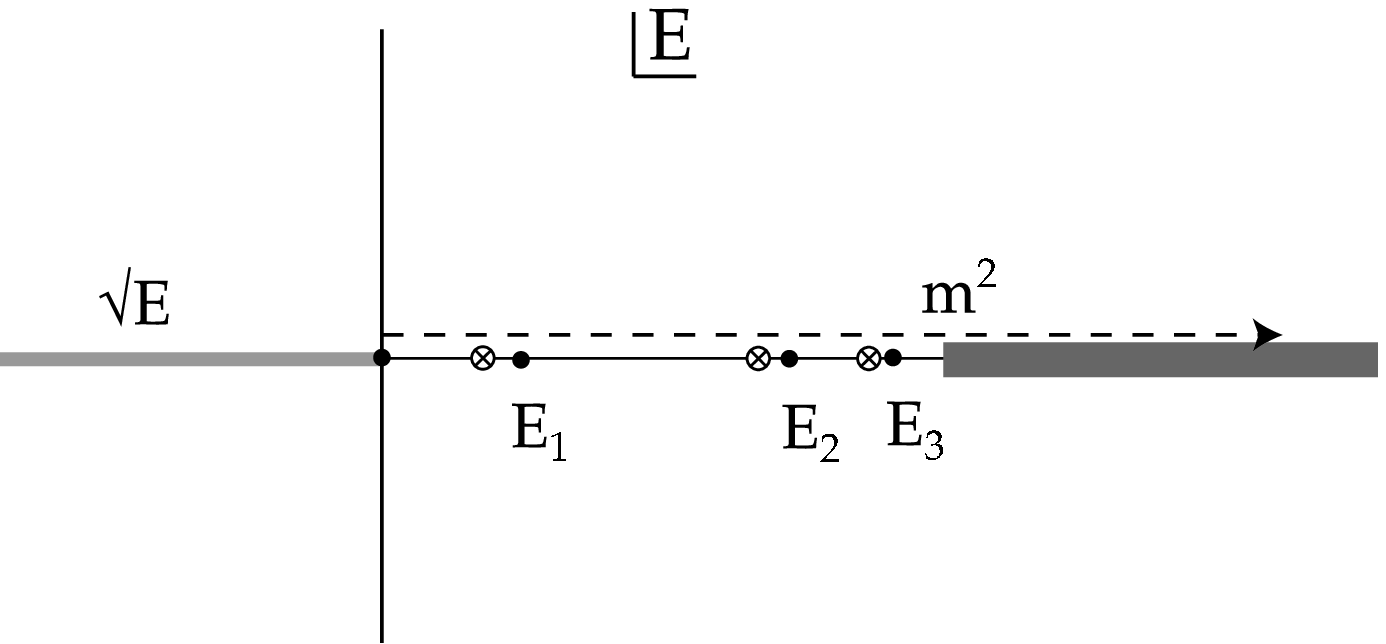}
\caption{\label{complexplane}\sl Analytic structure of the
integrand of eq.~(\ref{eq12.1}) in the complex $E$-plane.   The
left hand cut comes from the $\sqrt{E}$, the right hand cut
beginning at $E=m^{2}$ comes from scattering states.  The pole
contributions are marked by solid circles.  They lie just above
the bound states of $V(x)$, marked with ${\otimes}$.}}

The result,
\begin{equation}
    \cF =  \frac{1}{2\pi} \int_{0}^{\infty} dE\sqrt{E}
    \frac{\pp^{2}}{\pp E\pp a}\ln\left(1+\lambda\cG_{0}(a,a,-E)\right)\ ,
\label{eq12.2}
\end{equation}
is particularly useful because the argument of the logarithm is
positive definite. Eq.~(\ref{eq12.2}) can be integrated by parts
without contributions at the limits because the surface term at
$E\to \infty$ is independent of $a$ and the surface term at $E=0$
vanishes:
\begin{equation}
\label{eq:forceE}
    \cF(a,\lambda,V) =  -\frac{1}{4\pi} \int_{0}^{\infty}  {dE}\frac{1}{\sqrt{E}}
    \frac{\pp }{\pp a}\ln\left(1+\lambda\cG_{0}(a,a,-E)\right).
\label{12.21}
\end{equation}
Here we have restored some of the arguments on the function $\cF$
as a reminder of its important variation with position, $\lambda$,
and background field, $V$.  It will be sometimes convenient to
introduce the imaginary momentum $\kappa=\sqrt{E+m^{2}}$ (here $E$
is the dummy variable in (\ref{eq:forceE})) in the Euclidean
domain,
\begin{equation}
    \cF(a,\lambda,V) =  -\frac{1}{2\pi} \int_{m}^{\infty}  {d\kappa}\kappa\frac{1}{\sqrt{\kappa^{2}-m^{2}}}
    \frac{\pp }{\pp a}\ln\left(1+\lambda\cG_{0}(a,a,-\sqrt{\kappa^{2}-m^{2}})\right)\ ,
\label{12.22}
\end{equation}

This result can be rewritten usefully by introducing the Jost
solutions to the Schr\"odinger equation, $\cH\psi=-E\psi$ (notice
the minus sign in front of $E$, due to the Wick rotation), which
are defined by their behavior as $x\to\pm \infty$\cite{texts},
\begin{eqnarray}
    \lim_{x\to\infty}\psi^{+}(\kappa,x)e^{\kappa x}&=& 1\nonumber\\
    \lim_{x\to-\infty}\psi^{-}(\kappa,x)e^{-\kappa x}&=& 1
    \label{eq15}
\end{eqnarray}
The boundary conditions of eqs.~(\ref{eq15}) render the functions
$\psi^{\pm}$ analytic for ${\rm Re}\ \kappa>0$.   On the contour
$-\infty<E<0$ needed in eq.~(\ref{12.22}) the Jost solutions are
real and fall exponentially in the limits given in
eq.~(\ref{eq15}). The Greens function can be written in terms of
the Jost solutions,
\begin{equation}
\cG_{0}(x,y,-\sqrt{\kappa^{2}-m^{2}}) =
\frac{T(\kappa)}{2\kappa}\psi^{+}(\kappa,x_{>})\psi^{-}(\kappa,x_{<})
\end{equation}
where $x_{>}$ ($x_{<}$) is the greater (lesser) of $x$ or $y$, and
$T(\kappa)$ is the transmission coefficient.  Upon substituting
for $\cG_{0}$ in eq.~(\ref{12.22}), we obtain
\begin{equation}
\cF(a,\lambda,V)= -\frac{\hbar c}{2\pi} \int_{0}^{\infty}
 {d\kappa}\kappa\frac{1}{\sqrt{\kappa^{2}-m^{2}}}\frac{\pp}{\pp a}\ln\left(1+\frac{\lambda T(\kappa)}{2\kappa}
 \psi^{+}(\kappa,a)\psi^{-}(\kappa,a)\right)\nonumber\\
\label{eq16}
\end{equation}
In the last equation we have restored the factors of $\hbar$ and
$c$ ($E=\kappa^{2}-m^{2}$ has units of $1/{\rm length}^{2}$ in
these units).

\subsection{$\lambda\to \infty$:  the Dirichlet case}

When the strength, $\lambda$, of the sharp background field
becomes large compared to the eigenvalue, $\omega$, the mode of
the field with frequency $\omega$ vanishes at $x=a$.  If all the
modes that contribute to the buoyancy have eigenvalues below
$\lambda$, then effectively the field itself obeys  the boundary
condition $\phi(a,t)=0$.  As discussed in
Ref.~\cite{Graham:2003ib}, this description is acceptable if the
physical observable of interest remains becomes independent of
$\lambda$ in the limit.  In the case of buoyancy force in one
dimension the limit $\lambda\to\infty$ exists,
\begin{eqnarray}
\lim_{\lambda\to\infty}\cF(a,\lambda,V)\equiv\cF_{\infty}(a,V)&=&
-\frac{1}{2\pi} \int_{0}^{\infty}
\frac{dE}{\sqrt{E}}\frac{\pp}{\pp a}\ln \cG_{0}(a,a,-E)
\label{eq:FdirichletE}
\earr
which can be also written as
\barr
\cF_{\infty}(a,V)&=&-\frac{1}{4\pi} \int_{m}^{\infty} d\kappa
\kappa\frac{1}{\sqrt{\kappa^{2}-m^{2}}}\left(
\frac{1}{\psi^{+}(\kappa,a)}\frac{\pp\psi^{+}(\kappa,a)}{\p a} +
\frac{1}{\psi^{-}(\kappa,a)}\frac{\pp\psi^{-}(\kappa,a)}{\p
a}\right).
\label{eq17}
\end{eqnarray}

It is instructive to study the contribution to the Casimir force
of the bound states in the $\lambda\to\infty$ limit.\footnote{The
analysis for finite $\lambda$ is similar, but technically more
complicated. In practice, following the discussion in the previous
section, one has to find the poles of $1+\lambda\ \cG_0(a,a,E)$
say $E(a)$, which are also eigenvalues of $\cH$, and proceed as
before.} Returning to the non-rotated formula eq.~(\ref{eq12.1}),
taking $\lambda\to\infty$ and substituting the Jost solutions
representation for the Greens function, we have
\begin{equation}
 \cF_{\infty}(a,V)= -\frac{1}{2\pi}{\rm Im}\int_{0}^{\infty}
dE  \frac{1}{\sqrt{E}}\frac{\pp}{\pp a}\ln
\left(\psi^{+}(E,a)\psi^{-}(E,a)\right) .
\label{eq18}
\end{equation}
Remembering that Jost solutions are real for $E<m^{2}$, it is
clear that the only contributions in this range come from possible
zeros in $\psi^{\pm}(E,a)$, where the integrand picks up an
imaginary part because the integration contour goes above the pole
on the Re $E$ axis.  The locations of these zeros depends on $a$,
so we label them as $E_{j ^{\pm}}(a)$.  Near the $j^{\rm th}$ zero
in $\psi^{\pm}$,
\begin{equation}
\psi^{\pm}(E,a)=(E-E_{j^{\pm}}(a)+i\epsilon)\left.\frac{\pp
\psi^{\pm}(E,a)}{\pp  E}\right|_{E=E_{j^{\pm}}(a)} + \ldots
\label{eq18.1}
\end{equation}
where we have restored the $i\epsilon$ to make the nature of the
singularity clear.  Substituting into eq.~(\ref{eq18}) and
extracting the imaginary part, we obtain a very simple expression
for the bound state contribution to $\cF_{\infty}$,
\begin{equation}
\left.\cF_{\infty}(a,V)\right|_{\rm bound \ states}= -\frac{\hbar
c}{4}\sum_{j^\pm}\frac{1}{\sqrt{E_{j^{\pm}}(a)}}
 \frac{\pp E_{j^{\pm}}(a)}{\pp a}= -\frac{\pp}{\pp a}\left[\frac{\hbar c}{2}
 \sum_{j^\pm} \sqrt{E_{j^{\pm}}(a)}\right]
\label{eq19}
\end{equation}
Here we have used the fact that the various partial derivatives
are related by
\begin{equation}
d\psi = \frac{\pp \psi}{\pp E}dE+\frac{\pp \psi}{\pp a}da =0
\label{eq20}
\end{equation}
along the contour $\psi^{\pm}(E_{j^{\pm}}(a),a)=0$ in the $E,a$
plane.

The simple form of eq.~(\ref{eq19}) has a simple origin:  the
$\{\sqrt{E_{j^{\pm}}(a)}\}$ are the eigenenergies of $\phi$ in
$V(x)$ constrained to vanish at $x=a$, so the quantity in brackets
is the bound state contribution to the Casimir energy for a scalar
field in the potential $V(x)$ \emph{forced to vanish at $x=a$}.
This is an equivalent formulation of the Casimir buoyancy in the
Dirichlet ($\lambda\to\infty$) limit.  From this interpretation it
is clear that the number of terms in the sum in eq.~(\ref{eq19})
lies between $N-1$ and $N$, where $N$ is the number of bound
states in $V(x)$.

To complete this parametrization of the Casimir buoyancy we write
the continuum contribution as an integral over the scattering
momentum $k$, so another expression for the total buoyancy in the
$\lambda\to\infty$ limit, equivalent to eq.~(\ref{eq17}) is
\begin{eqnarray}
\cF_{\infty}(a,V)=-\frac{\pp}{\pp a}\left[\frac{\hbar c}{2}
 \sum_{j^\pm} \sqrt{E_{j^{\pm}}(a)}-\frac{\hbar c}{2\pi}{\rm Im}\int_{0}^{\infty}dk
\frac{k}{\sqrt{k^{2}+m^{2}}}\ln\left(\psi^{+}(E,a)\psi^{-}(E,a)\right)
\right]
\label{eq21}
\end{eqnarray}
If the $k$-integration were rotated to the positive imaginary
axis, $k\to i\kappa$, two kinds of contributions would arise: a)
from the cut from $\frac{1}{\sqrt{k^{2}+m^{2}}}$ and b) from poles
in $\psi^{+}\psi^{-}$ corresponding to the same states counted in
the sum over $j$.  The pole contribution would exactly cancel the
sum and the integral over $\kappa$ would yield eq.~(\ref{eq17}) as
it must.

\section{Approximations and Special Cases}

Casimir buoyancy is an unfamiliar phenomenon.  Its properties are
not readily apparent from the general expressions,
eqs.~(\ref{eq17}) and (\ref{eq21}).  In this section we study the
buoyancy in special situations where it simplifies.  First we look
at smooth, deep potentials where the WKB approximation applies.
Next we look at reflectionless potentials, where we show that only
the bound states contribute to the buoyancy.  Third, we look at
the buoyancy in the domain beyond the range of $V(x)$, and finally
we study the first Born approximation, where the buoyancy is
simple, but not local.

\subsection{WKB approximation}

If the potential is smooth and strong, a WKB expansion can be made
for the Casimir buoyancy.  We work with the Wick rotated form,
eq.~(\ref{eq17}).  To apply WKB \cite{landau} we introduce a
fictitious ``$\tilde\hbar$'' into the Hamiltonian, $\cH \to
-\tilde\hbar^{2}\frac{d^{2}}{dx^{2}}+W(x)$ (henceforth we will
write $W(x)$ for $m^2+V(x)$) and write
\begin{equation}
\psi^{\pm}(E,x)=\exp{\left(-\frac{1}{\tilde\hbar}s_{0}(x,E)+s_{1}(x,E)+...\right)}.
\label{eq22}
\end{equation}
and find
\begin{eqnarray}
s_{0}(x,E)&=&\pm\int_0^{x}dy\sqrt{-E+W(y)}\nonumber\\
s_{1}(x,E)&=&-\frac{1}{4}\ln(-E+W(x))
\label{eq23}
\end{eqnarray}
so the domain $E<0$ corresponds to the WKB forbidden region (since
$W(x)>0$, which we have assumed from the outset) and gives real
$s_{0,1,...}$, $\psi_\pm$ and $\cG_0$.

The criterion for the validity of the WKB expansion (with
$\tilde\hbar$ set to unity), $s_{1}'\ll s_{0}'$, reduces to
\begin{equation}
\frac{d}{dx}\frac{1}{\sqrt{W(x)}}\ll 1,
\label{eq24}
\end{equation}
from which we conclude that $W'(x)/W^{3/2}$ should be small and in
particular  where  $W=m^{2}+V(x)$  should not vanish\footnote{If
$W\sim x^{2n}$ then $W'/W^{3/2}\sim x^{-n-1}$ becomes arbitrarily
large as $x\to 0$.} (\emph{i.e.} at turning points).  Hence  the
WKB approximation applies to potentials which are ``deep'' in the
sense that $\int dy \sqrt{W(y)}\gg 1$ and far from turning points
where $W(x)=0$.

It is straightforward to construct the Greens function, in the WKB
approximation,
\begin{equation}
\label{eq:GWKB}
\cG_{0\ {\rm WKB}}(a,a,-E)=\frac{1}{2\sqrt{E+W(a)}}
\label{eq25}
\end{equation}
and, upon substituting into eq.~(\ref{eq:forceE}), and performing
the integral over $E$, we find
\barr
\cF(\lambda,a)_{\rm WKB} &=&  \frac{\pp}{\pp a}\frac{\hbar
c}{4\pi}\left( \pi\sqrt{W(a)}+ \lambda\ln(W(a))-\sqrt{\lambda^{2}
-4W(a)}\ln\frac{\lambda-\sqrt{\lambda^{2}-4W(a)}}{\lambda+\sqrt{\lambda^{2}-4W(a)}}
\right)\nonumber\\
&\equiv&-\frac{\p }{\p a}\Omega(a,\lambda)_{\rm WKB}.
\label{eq26}
\earr
\textit{is a local function of the potential, $V(a)$}. We call
this function the \emph{quantum potential} $\Omega(a,\lambda)$.

Equation (\ref{eq26}) simplifies considerably in the Dirichlet
limit,
\begin{equation}
\label{eq:WKB0}
\cF_{\infty\ {\rm WKB}}(a)=\frac{\pp}{\pp a}\frac{\hbar
c}{4}\sqrt{m^{2}+ V(a)}\equiv-\frac{\p}{\p a}\Omega(a)_{\rm WKB}.
\end{equation}
A remarkably simple result for such a complex phenomenon. One
general feature can be easily deduced from this form of the
quantum potential $\Omega$: it decreases when the potential $V(x)$
increases and hence the force felt by the plate is a
\emph{buoyancy} force. We will encounter this phenomenon
throughout the rest of this paper in extensions (to regions where
the simple WKB form is not valid, to non-zero temperature and to
higher number of dimensions) and several examples.

\subsection{Points at which $V(x)+m^2=0$}
\label{sec:ho}

When the potential is smooth (with a definite length-scale $b\sim
[\frac{1}{W}\frac{dW}{dx}]^{-1} $)  but there exists an $x_0$ such
that $ V(x_0)+m^2 =0$ (or in general an interval  where
$(V(x)+m^2)b^2\ll 1$) then WKB cannot be applied. The simple
formula eq.~(\ref{eq:WKB0}) breaks down. For example in the case
of symmetric potential $m^2+V(x)=x^2$ the force must vanish at
$x=0$ for symmetry, while eq.\ (\ref{eq:WKB0}) predicts $\cF\to
{\rm constant}$.

To understand what really happens when the delta function reaches
the point $x_0$ one needs a more clever guess than WKB. Moreover,
adding more terms of the WKB series cannot help since the
asymptotic nature of WKB means that adding terms improves the
result for \emph{larger} values of $b^2(V+m^2)$, so certainly not
close to $x_0$. The formally correct procedure would be to find a
differential equation `similar' to the one we are studying but
solvable and a smooth map from one to the other \cite{BerryMount}.
In this way we would obtain a \emph{uniform} approximation near
the turning point $x_0$. However such an analysis goes beyond the
goal of this paper, so to clarify the situation we will assume
that close to $x_0=0$ we have (remember that $W(x)\geq 0$)
\beq
W(x)=V(x)+m^2\simeq \omega^2 x^2+\Ord{x^3}.
\eeq
If we have $0\neq W(0)\equiv W_0\ll 1/b^2$, this constant could be
reabsorbed by a shift of $E$ and the range of integration in
eq.~(\ref{eq20}), we will consider this case at the end of this
section.

We then study the propagator in the neighborhood of $x_0=0$
\emph{i.e.}\ the propagator $\cG_0$ of the equation
\begin{equation}
-\phi''(x)+\omega^2x^2\phi(x)=E\phi(x)
\end{equation}
which is the usual harmonic oscillator problem. We now set
$\omega=1$ and we will reintroduce it only at the end of the
calculation to have dimensionally correct results.

From the two independent Jost solutions of this equation it is
straightforward to write the the propagator for the harmonic
oscillator as
\begin{equation}
\label{eq:Gharm}
\cG_{0}(a,a,E)= \frac{2^{-(E+1)/2}}{\sqrt{\pi}}\Gamma
\left(\frac{1-E}{2}\right)e^{- a^2}H_{\frac{E-1}{2}}(
a)H_{\frac{E-1}{2}}(- a)
\end{equation}
where $H$ is the Hermite function, a generalization of the Hermite
polynomials related to the parabolic cylinder function
\cite{Abramowitz}. Notice that the poles of the gamma function
give the correct spectrum of the bound states.

One can then calculate the buoyancy force for any $\lambda$, by
inserting eq.~(\ref{eq:Gharm}) into eq.~(\ref{eq:forceE}) and
performing the integral numerically. In the Dirichlet case
$\lambda\to\infty$ however the results simplifies using
eq.~(\ref{eq:FdirichletE}) to
\begin{equation}
\cF_{\infty\ {\rm h.o.}}=-\frac{1}{4\pi}
\int_0^\infty\frac{dE}{\sqrt{E}}\left((1+E)\bigg(
\frac{H_{-(3+E)/2}(-a)}{H_{-(1+E)/2}(-a)}-\frac{H_{-(3+E)/2}(a)}{H_{-(1+E)/2}(a)}\bigg)-2a\right).
\end{equation}
In this case the expression for small $a$ can be recovered by
expanding the integrand in powers of $a$ and integrating term by
term (the expansion can be carried on to any order of $a$  because
the integrals over $E$ converge):
\begin{eqnarray}
\cF_{\infty\ {\rm
h.o.}}&\simeq&\frac{1}{4\pi}\left(a\left(\int_0^\infty
dE\frac{1}{\sqrt{E}}
    \left(
      \frac{8\,
         {\Mfunction{\Gamma}(
           \frac{3 + E}{4})}^2}{
           {\Mfunction{\Gamma}(
           \frac{1 + E}{4})}^2}
      -2E\right)\right)+\Ord{a^3}\right)\nonumber\\
      \label{eq:Harmexpan}
      &\simeq&\frac{1}{4\pi}\omega^{3/2}a(3.24...)+\Ord{\omega^{5/2} a^3}
\end{eqnarray}
which again exhibits the buoyancy phenomenon, \emph{ie.} the
Casimir force pushes the plate toward points at higher potential.
Notice that the force vanishes as $a\to x_0=0$ as it should from
symmetry arguments.\footnote{If we want to compare this result
with an exact one we can choose $n=4$ Poschl-Teller (see Section
\ref{sec:PT}) with $m=\sqrt{20}$ for which $V(x)+m^2\sim 20 x^2$
near $x=0$. In this case $\cF_{\rm exact}/\cF_{\rm approx}=0.94$
as $a\ll 1$. Such an agreement must be considered impressive,
since the propagator is not a local function of the potential $V$
and a local approximation to $V$ does not guarantee at all a local
approximation to $G$. Evidently, for sufficiently smooth
potentials this turns out to be the case.}

If $V(x_0)+m^2=W_0$ and $W_0\ll 1/b^2$ ($b^2= 1/\omega$ in this
case) one can repeat the preceding derivation, change the lower
limit of integration over $E$ from $0$ to $W_0b^2$ ($W_0/\omega$
in the case at hand) and write
\beq
\label{eq:fho}
\cF_{\infty\ {\rm h.o.}}\simeq\frac{1}{4\pi}\omega^{3/2}a
f(W_0/\omega),
\eeq
where we have defined
\beq
f(x)=\int_x^\infty dy\frac{1}{\sqrt{y-x}}
    \left(
      \frac{8\,
         {\Mfunction{\Gamma}(
           \frac{3 + y}{4})}^2}{
           {\Mfunction{\Gamma}(
           \frac{1 + y}{4})}^2}-2y\right),
\eeq
and $f(0)=3.24...\ $ . Although  we do not have an analytic
expression for $f$ it can be studied numerically with ease and to
a great accuracy.  In the opposite limit $W_0/\omega\gg 1$ one can
still shift the range of integration and then, by working out the
asymptotic limit of the propagator eq.~(\ref{eq:Gharm}) as
$E/\omega\gg 1$, one finds the now familiar WKB result
eq.~(\ref{eq:WKB0}). We will apply eq.\ (\ref{eq:fho}) to the case
of P\"oschl-Teller potential in Section \ref{sec:PT}.

One can also study higher order zeros (or minima) of $W$,
\emph{i.e.}\ points where $W(x)\sim x^{2n}$ in a similar fashion.
The Jost solutions of such a problem should be found, possibly in
a series for small $x$, and by means of them the propagator can be
written explicitly in the neighborhood of the minimum $x=0$. Here
we will not extend our analysis to those cases since no
qualitatively new phenomenon arises.

\subsection{Temperature dependence in the WKB approximation}

One can also study the case where the field is in a non-zero
temperature thermal state rather than a vacuum state. The Casimir
energy is then the integral of the energy momentum tensor
component $T_{00}$ evaluated on the thermal state. The result is
\cite{optical2}
\beq
\label{eq:temp1}
\cE_\lambda=-\frac{1}{2\pi}\Im\int_0^\Omega
dE\sqrt{E}\coth\left(\beta\frac{\sqrt{E}}{2}\right)\frac{\p}{\p
E}\log\left(1+\lambda\cG_0(a,a,E)\right).
\eeq
One can Wick-rotate this expression as well, since the Matsubara
poles are on the negative real $E$ axis. After the rotation the
Greens function becomes real (the $i$ from $\sqrt{E}$ cancels with
an $i$ from $\coth$) and one has an imaginary part from the $\cot$
function. Dropping a term $\propto \lambda\log\Omega$ arising from
the semicircle at $|E|=\Omega$ one has
\beq
\cE_\lambda=-\frac{1}{2\pi}\int_0^\Omega dE\sqrt{E}\ \Im\Big\{
\cot\left(\beta\frac{\sqrt{E}}{2}-i\epsilon\right)\Big\}\frac{\p}{\p
E}\log\left(1+\lambda\cG_0(a,a,-E)\right).
\eeq
In the limit $\epsilon\to 0$ one has
$\Im\cot(x-i\epsilon)=\pi\sum_n\delta(x-n\pi)$.
The final expression is then a sum over Matsubara frequencies.
\beq
\cE_\lambda=-\frac{1}{2}\int_0^\Omega
dE\sqrt{E}\sum_n\delta\left(\beta\frac{\sqrt{E}}{2}-n\pi\right)
\frac{\p}{\p E}\log\left(1+\lambda\cG_0(a,a,-E)\right).
\eeq

The study of this expression would require a paper on its own.
Here we will limit our analysis of the WKB case, since a closed
exact expression can be easily obtained in this case. To obtain it
however, we have to go back to the original, non Wick-rotated WKB
expression eq.~(\ref{eq:temp1}) (we again use the notation
$W(x)=V(x)+m^2>0$)
\beq
\cE_\lambda\simeq-\frac{1}{2\pi}\Im\int_0^\Omega
dE\sqrt{E}\coth\left(\beta\frac{\sqrt{E}}{2}\right)\frac{\p}{\p
E}\log\left(1+\frac{i\lambda}{2\sqrt{E+i\epsilon-W(a)}}\right).
\eeq

The force in the Dirichlet limit (the cutoff $\Omega$ can again be
sent to $\infty$ without problems) is
\beq
\cF_\infty\simeq\frac{1}{4\pi}\frac{dV}{da}\Im\int_0^\infty
dE\sqrt{E}\coth\left(\beta\frac{\sqrt{E}}{2}\right)\frac{\p}{\p
E}\left(\frac{1}{E-W(a)+i\epsilon}\right),
\eeq
and by trading the $E$ derivative with a $W(a)$ derivative inside
the integral, taking the imaginary part and the limit $\epsilon\to
0$ we find
\begin{eqnarray}
\cF_\infty(a,\beta)&\simeq&-\frac{1}{4\pi}\frac{dW}{da}\frac{\p}{\p
W}\int_0^\infty dE\sqrt{E}\coth\left(\beta\frac{\sqrt{E}}{2}\right)(-\pi\delta(E-W))\nonumber\\
&=&\frac{1}{4}\frac{d}{da}\left(\sqrt{W}\coth(\beta
\sqrt{W}/2)\right)\equiv-\frac{d}{da}\Omega(a,\beta),
\end{eqnarray}
where the last expression defines again a local \emph{thermal
quantum potential}
\beq
\Omega(a,\beta)=-\frac{1}{4}\sqrt{W}\coth(\beta\sqrt{W}/2).
\eeq

In the low temperature limit $T\to 0$ we have $\Omega\to
-\sqrt{W}/4$ as it should (compare with eq.~(\ref{eq:WKB0})). In
the high temperature limit $T\to\infty$, $\beta\to 0$ one finds
$\cF_\infty\propto T^{-1} dW/da$. Contrary to what one could
expect from the known `classical limit' $\cF\propto T/a$ of
Casimir force between rigid bodies \cite{classicaltemp, optical2},
the force goes to zero when the temperature grows indefinitely.
However here we are in a totally different limit, where the
background potential is slowly varying and we are considering only
the zero reflection term for $\cG_0$. The rigid bodies expansion
of $\cG$ that gives rise to the well-known Casimir force actually
is made up of non-local reflection contributions \cite{optical1}.

This expression is valid whenever WKB is valid, hence when the
length-scale of the potential $b$ is such that $W b^2\gg 1$. We
have already discussed the buoyancy effect, \emph{i.e.} the
quantum potential has a maximum where $V$ has a minimum. A
non-zero temperature does not modify this prediction qualitatively
as can be seen from Figure \ref{temperature}.

\FIGURE{
\includegraphics[width=12cm]{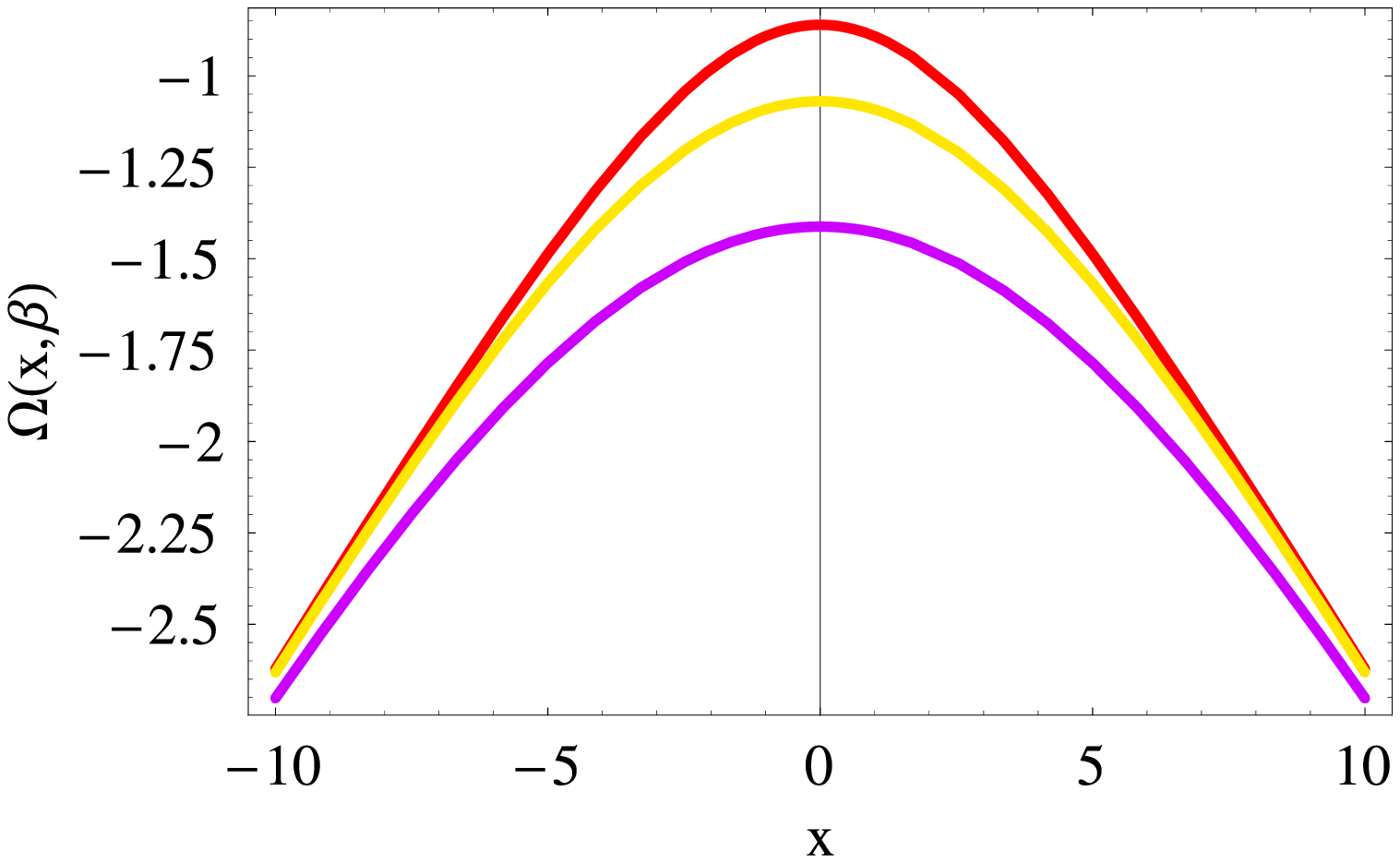}
\caption{\label{temperature}\sl Turning on the temperature does
not affect the Casimir buoyancy qualitatively.  Here we plot
$\Omega$ for the potential $W(x)=10+x^2$ and $\beta=1,0.6,0.4$
from up down respectively. The minimum of $W$ at $x=0$ is always a
maximum of $\Omega_{\rm WKB}$. }}

\subsection{Reflectionless potentials}

The continuum contribution to the Casimir buoyancy vanishes in the
Dirichlet limit if a potential is reflectionless.  For a
reflectionless potential, $\psi^{+}(k,x)\to \frac{1}{T(k)}e^{ikx}$
as $ x\to-\infty$, so
$\psi^{-}(k,x)=\frac{1}{T^{*}(k)}\psi^{+*}(k,x)$.  In this case
$\psi^{+}(k,a)\psi^{-}(k,a)=|\psi^{+}(k,a)|^{2}/T^{*}(k)$. Looking
back to eq.~(\ref{eq21}), we see that the imaginary part of the
logarithm is independent of $a$ so the force vanishes.  Thus the
Casimir buoyancy in a reflectionless potential is entirely
determined by the bound states.

\subsection{Beyond the range of the potential}

Here we consider a background potential which vanishes identically
for $|x|>b$.  The same results apply to a short range potential
($V(x)\propto e^{-\mu|x|}$) in the limit $x\to\infty$.  For $x<-b$
in  the Euclidean domain,
\begin{eqnarray}
\psi^{-}(\kappa,x)&=& e^{ \kappa x}\nonumber\\
\psi^{+}(\kappa,x)&=&\frac{1}{T(\kappa)}e^{-\kappa x}+\frac{
R(\kappa)}{T(\kappa)}e^{\kappa x}
\label{eq28}
\end{eqnarray}
 The Greens function becomes
\begin{equation}
\cG_{0}(a,a,-\sqrt{\kappa^{2}-m^{2}})=\frac{1}{2\kappa}\left(1+
R(\kappa)e^{ 2\kappa a}\right)
\label{eq29}
\end{equation}
so the buoyancy can be written,
\begin{equation}
\cF(\lambda,a)=-\frac{\hbar c}{2\pi}\int_{m}^{\infty}\frac{\kappa
d\kappa}{\sqrt{\kappa^{2}-m^{2}}}\frac{\pp}{\pp a}\ln\left(
1+\frac{\lambda}{2\kappa}\left(1+  R(\kappa)e^{ 2\kappa
a}\right)\right)\quad\hbox{for}\ a<-b.
\label{eq30}
\end{equation}
For $x>b$ a similar expression holds with $ R\to \overline R$, the
reflection coefficient on the right. If the potential is
symmetric, then $\overline R=R$. If the mass, $m$, of the
fluctuating field is non-zero, then $\cF$ falls like $e^{-2ma}$
when $a>b$.  The explicit form depends on the reflection
coefficient in the Euclidean domain.  In Section IV the explicit
example of a $\delta$-function background is studied.

\subsection{First Born approximation}

When the background potential is weak the Casimir buoyancy can be
expanded in powers of $V$.  The first term is simple  when $m=0$
and $\lambda\to\infty$.  A straightforward calculation gives the
Greens function to ${\cal O}(V)$,
\begin{equation}
\cG_{0}(a,a,E)=\frac{i}{2k}+\frac{1}{4k^{2}}\left(e^{2ika}\int_{-\infty}^{a}dye^{-2iky}V(y)+e^{-2ika}\int_{a}^{\infty}dye^{2iky}V(y)
\right)+ {\cal O}(V^{2}).
\label{eq31}
\end{equation}
$\cG_{0}$ has no bound states at this order, so only the continuum
contribution in eq.~(\ref{eq21}) need be calculated.
Straightforward evaluation leads to
\begin{equation}
\cF_{1}= -\frac{\hbar
c}{2\pi}\int_{-\infty}^{\infty}\!\!\!\!\!\!\!\!\!\!\!\! -\qquad
\frac{dz}{2z} V(z+a)=-\frac{\hbar c}{4}{\mathbf H}[V,a]
\label{eq32}
\end{equation}
(where the integral is intended as the Cauchy principal value)
which is a non-local functional of $V$ known to mathematicians as
the \emph{Hilbert transform} ${\mathbf H}$ of $V$
\cite{Hilberttransf} at the point $a$.

\section{Examples}

In this section we report the Casimir buoyancy in three explicit
sample background potentials and use them to study the domains of
application of the approximations in the previous section.  First
we treat the potential $\ell(\ell+1)/x^{2}$ on the half-line
$x>0$.  Second we explore the family of P\"oschl-Teller
potentials, $-n(n+1) {\rm sech}^{2} x $.  Finally we study the
$\delta$-function, $\beta\delta(x)$, which has been studied in
other contexts\cite{many}.

\subsection{$V(x)=\ell(\ell+1)/x^{2}$}

The potential $V_{0}/x^{2}$ leads to a well defined problem on the
half-line $x>0$ when $V_{0}$ is positive.  It is convenient to
parameterize $V_{0}$ by $\ell(\ell+1)$, so we can use  results
familiar from the study of three dimensional central potentials.
Note, however, that $\ell$ need not be integer. Since $V(x)$ is
positive definite there is no obstruction to taking $m=0$, which
we adopt to simplify the calculation. The formalism of Section II
has to be modified slightly on account of the boundary at $x=0$.
In particular, the Jost solution $\psi^{-}(x,E)$ has to be
replaced by the solution, $\phi(x,E)\propto j_{\ell}(kx)$,
\textit{regular at $x=0$}.  The other Jost solution is given by
$h^{(1)}_{\ell}(kr)$.   With this adaptation, and noting that the
potential has no bound states, we can compute the Casimir buoyancy
as an integral over scattering states, rotated to imaginary
momentum,
\begin{equation}
\cF(\lambda,\ell,a)=-\frac{\hbar c}{2\pi
a^{2}}\int_{0}^{\infty}d\xi \left(-\ln\left(1+\lambda
aI_{\tilde\ell }(\xi)K_{\tilde\ell}(\xi)\right)+\frac{\lambda a
I_{\tilde\ell}(\xi) K_{\tilde\ell}(\xi)}{1+\lambda
aI_{\tilde\ell}(\xi)K_{\tilde\ell}(\xi)}\right)
\label{eq33}
\end{equation}
where $\tilde\ell=\ell+1/2$, and $I_{\nu}$ and $K_{\nu}$ are
modified Bessel functions.  It is easy to verify that the $\xi$
integral is convergent. In the Dirichlet limit the buoyancy is
simply proportional to $1/a^{2}$ (for dimensional reasons it could
not be different),
\begin{equation}
\lim_{\lambda\to\infty}\cF(\lambda,\ell,a)=\cF_{\infty}(\ell,a)
=-\frac{\hbar c}{2\pi a^{2}}\int_{0}^{\infty}d\xi\left(
1+\xi\frac{I_{\tilde\ell}'(\xi)}{I_{\tilde\ell}(\xi)}+
\xi\frac{K_{\tilde\ell}'(\xi)}{K_{\tilde\ell} (\xi)}\right)
\label{eq34}
\end{equation}
\FIGURE{
\includegraphics[width=12cm]{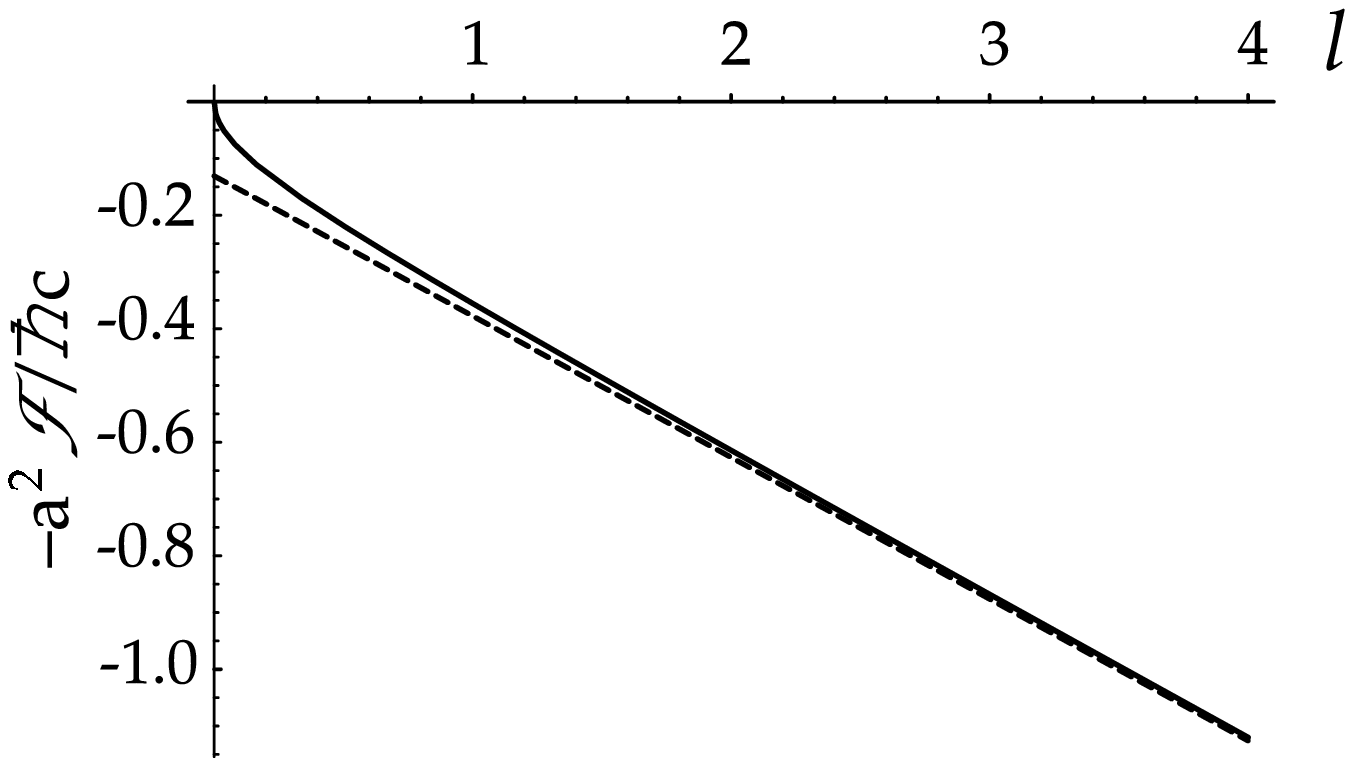}
\caption{\label{ellinfinity}\sl Casimir buoyancy, $-a^{2}\cF/\hbar
c$, for $V(x)=\ell(\ell+1)/x^{2}$ in the Dirichlet limit.  The
dashed curve is the exact result of eq.~(\ref{eq34}).  The solid
curve is the WKB approximation, eq.~(\ref{eq35}). }}
\FIGURE{
\includegraphics[width=12cm]{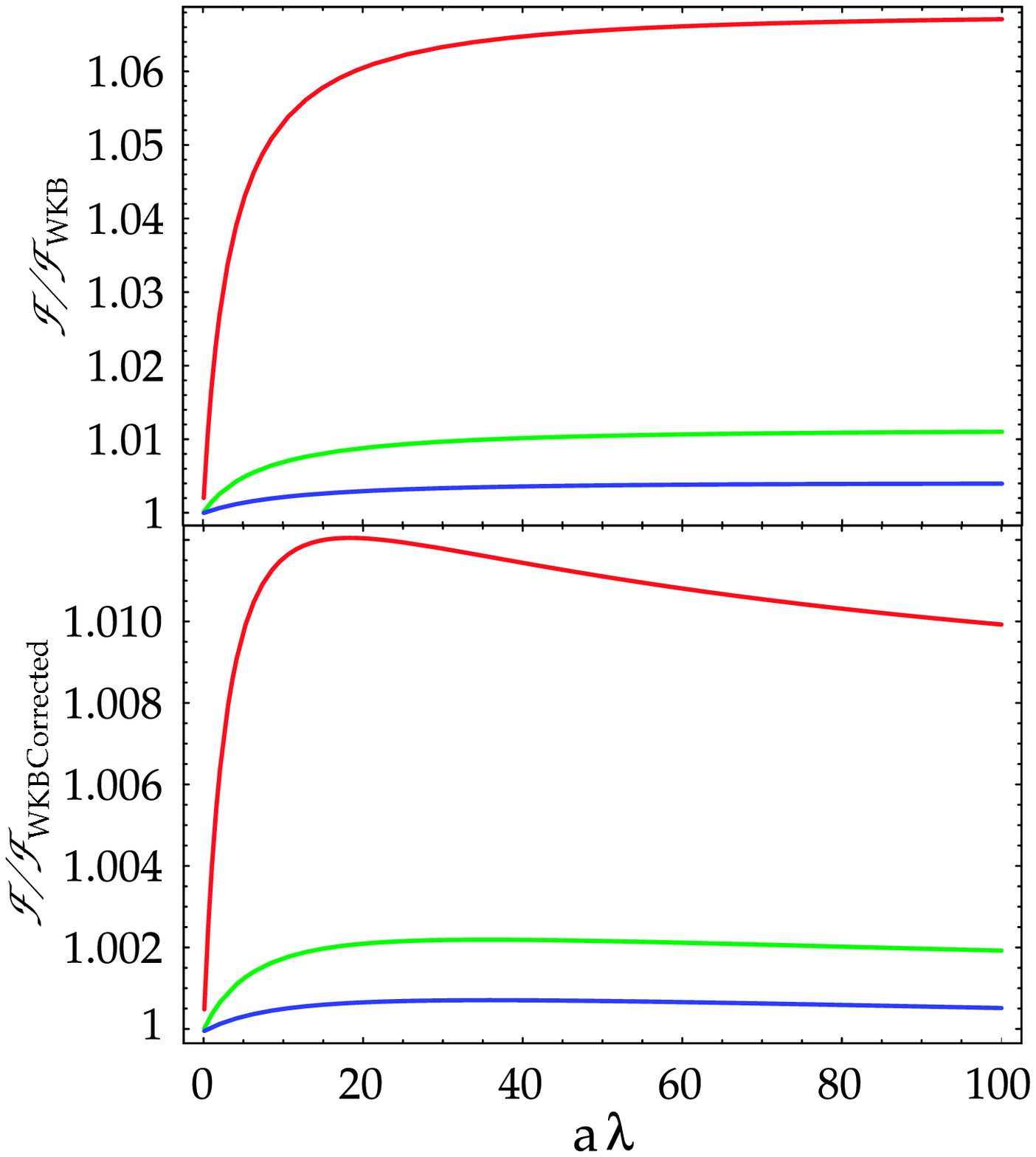}
\caption{\label{fig:lwkb}\sl (a) Ratio of Casimir buoyancy to the
WKB approximation for $V(x)=\ell(\ell+1)/x^{2}$. Up to down (red
to blue) $\ell=1,3,5$; (b) Same as (a) with the ``correction'',
$\ell(\ell+1)\to (\ell+\half)^{2}$. }}

The potential $\ell(\ell+1)/x^{2}$ is neither weak nor
reflectionless.  It is however smooth and deep, so the WKB
approximation should be accurate.  $\cF_{\rm WKB}(\lambda,\ell,a)$
can be calculated easily from eq.~(\ref{eq:WKB0}), and the limit
$\lambda\to\infty$ is particularly simple,
\begin{equation}
\cF_{{\rm WKB}\ \infty}( \ell,a) = -\frac{\hbar
c\sqrt{\ell(\ell+1)}}{4a^{2}}
\label{eq35}
\end{equation}
The exact result and the WKB approximation are compared in the
$\lambda\to\infty$ limit in Fig.\ref{ellinfinity}.  The difference
is very small once $\ell\gtrsim 1$.  In fact WKB does even better:
if we make the standard Langer replacement,
$\sqrt{\ell(\ell+1)}\to (\ell+\half)$\cite{landau, BerryMount},
then the WKB and exact results coincide within the widths of the
lines in Fig.~\ref{ellinfinity}.

In Fig.~\ref{fig:lwkb} we compare the exact result with the WKB
approximation for finite $\lambda$. Clearly WKB is an excellent
approximation in this case.

\subsection{P\"oschl-Teller potentials}
\label{sec:PT}
The potentials $V_{n}^{\rm PT}(x)=-n(n+1){\rm sech}^{2}(x)$,
$n=1,2,3,...$, are reflectionless and the associated solutions to
the Schr\"odinger equation can be expressed in terms of elementary
functions.\footnote{The apparent dimensional inconsistency in
$V^{\rm PT}_{n}$ needs a word of explanation.  If we begin with a
dimensionally correct Hamiltonian, $-d^{2}/dx^{2}+m^{2}-V_{0}{\rm
sech}^{2}(x/b)$ and define a dimensionless unit of distance,
$z=x/b$, then $V_{0}b^{2}$ is the dimensionless potential which
equals $n(n+1)$ in the P\"oschl-Teller problem.}  The Casimir
buoyancy can be computed from the bound states alone.  For
simplicity we restrict ourselves to the $\lambda\to\infty$ limit
in this case. In practice, it is easier to compute the integrals
in the Euclidean region using eq.~(\ref{eq17}) since the
scattering wavefunctions, $\psi^{\pm}(E,a)$ are easy to construct
but the roots of $\psi^{\pm}(E_{j^{\pm}}(a),a)=0$ are hard to find
for $n>2$.  The necessary Jost solutions for imaginary momentum
are given by
\begin{equation}
\psi^{\pm}_{n}(\kappa,x)=\left.\left(\frac{d}{dx}-\tanh x
\right)\left( \frac{d}{dx}-2\tanh x
\right)...\left(\frac{d}{dx}-n\tanh x \right)
e^{ikx}\right|_{k=\pm i\kappa}
\label{eq36}
\end{equation}
up to an inessential normalization.  The necessary product of Jost
solutions is given by
\begin{eqnarray}
\psi^{+}_{1}(\kappa,a)\psi^{-}_{1}(\kappa,a)&=& \tanh^{2}a-\kappa^{2} \nonumber\\
\psi^{+}_{2}(\kappa,a)\psi^{-}_{2}(\kappa,a)&=&
(1-\kappa^{2}-3\kappa\tanh a
-3\tanh^{2}a)(1-\kappa^{2}+3\kappa\tanh a -3\tanh^{2}a)
\label{eq37}
\end{eqnarray}
for example for $n=1$ and $n=2$.  The required integrals can be
performed explicitly for $n=1$ and numerically for larger $n$.

The case $n=1$ is particularly simple and instructive:
\begin{equation}
\cF^{\rm PT}_{1}(m,a)=\hbar c\frac{\tanh a\ {\rm
sech}^{2}a}{2\sqrt{ m^{2}-\tanh^{2}a}}
\label{eq38}
\end{equation}
As  usual the Casimir force is in the direction opposite to the
gradient of $V(x)$.  The WKB approximation to the P\"oschl-Teller
potentials is
\begin{equation}
\cF^{\rm PT}_{n\ {\rm WKB}}(m,a)=\hbar c n(n+1)\frac{\tanh a \
{\rm sech}^{2}a}{4\sqrt{m^{2}-n(n+1){\rm sech}^{2}a}}
\label{eq39}
\end{equation}
In order to understand how well the WKB result approximates the
exact results we have plotted the force $\cF$ divided by
$\sqrt{n(n+1)}$ for fixed $\mu\equiv m/\sqrt{n(n+1)}=\sqrt{2}$. In
these variables, the WKB result eq.~(\ref{eq39}) is independent of
$n$ and the the exact curves tend to the WKB curve as $n$ is
increased. This is no surprise since for large $n$ the
Poschl-Teller potential becomes more and more semiclassical.

Even for large $n$, however, if $\mu\simeq 1$ we cannot use WKB
since the minimum of the potential at $x=0$ is \emph{almost} a
turning point, \emph{i.e.}\ $(V(0)+m^2)b^2\ll 1$ (remember we set
$b=1$). In this case however we can resort to the harmonic
oscillator approximation of Section \ref{sec:ho} which, for $a\to
0$ and specializing to the Poschl-Teller potential, takes the form
(the notation is the same as in Section \ref{sec:ho})
\beq
\cF_{n\ {\rm h.o.}}^{\rm PT}\simeq \frac{1}{4\pi}(n(n+1))^{3/4}a
f\left(\frac{m^2-n(n+1)}{\sqrt{n(n+1)}}\right). \label{eq:ptho}
\eeq
The results are plotted in Figure \ref{fig:ptho}. As in the
comparison with WKB the agreement is better the higher $n$,
\emph{i.e.}\ the more `semiclassical' is the potential.

\FIGURE{\includegraphics[width=12cm]{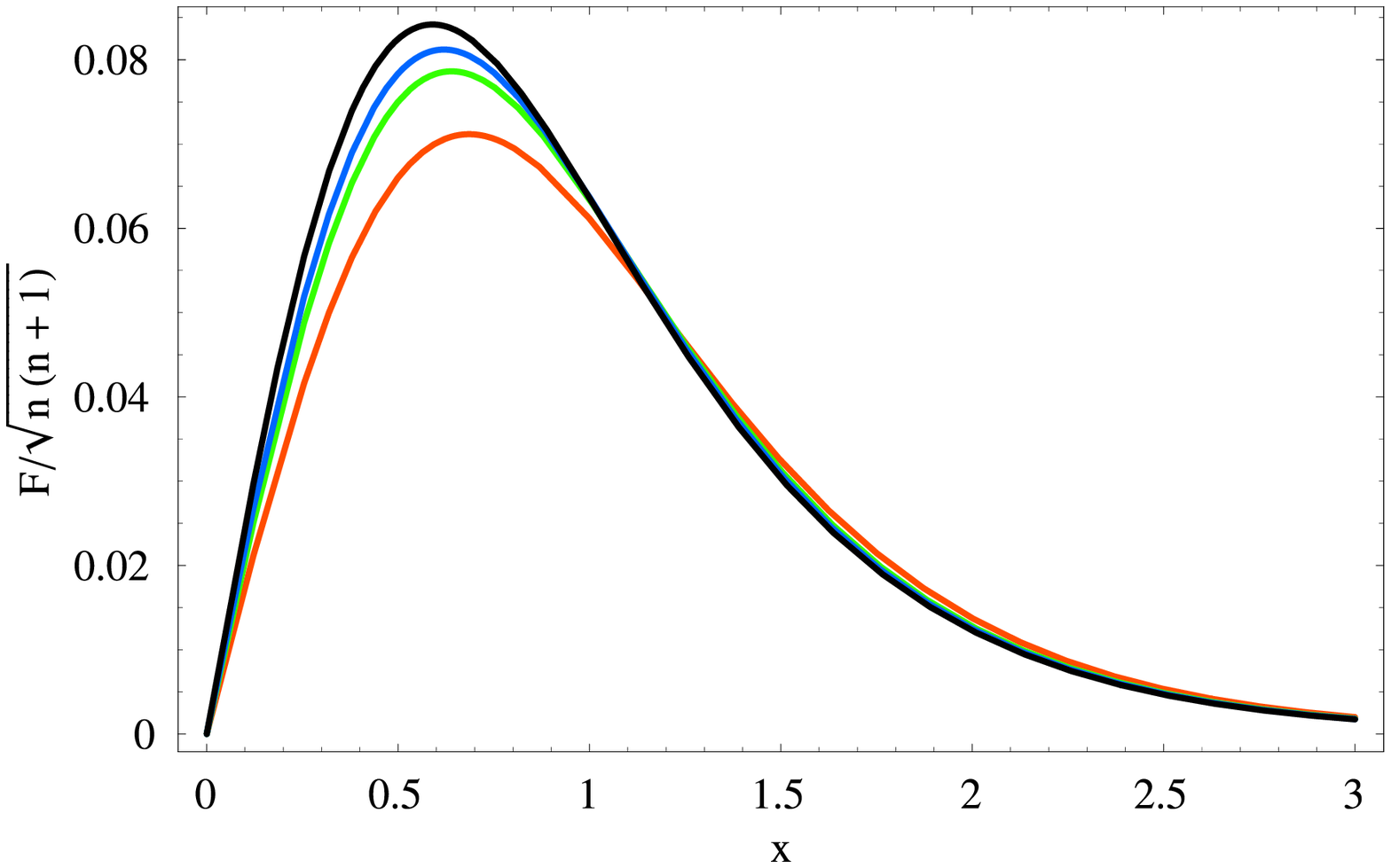}
\caption{\label{fig:pt}\sl Casimir force for Poschl-Teller
potential in the Dirichlet limit. Exact vs.\ WKB results. From
down up (red, green and blue) we have the exact results for
$n=1,2,3$ and in on the top, in black, the WKB approximation.
$\mu^2\equiv m^2/n(n+1)=2$ for all the curves.}}

\FIGURE{
\includegraphics[width=12cm]{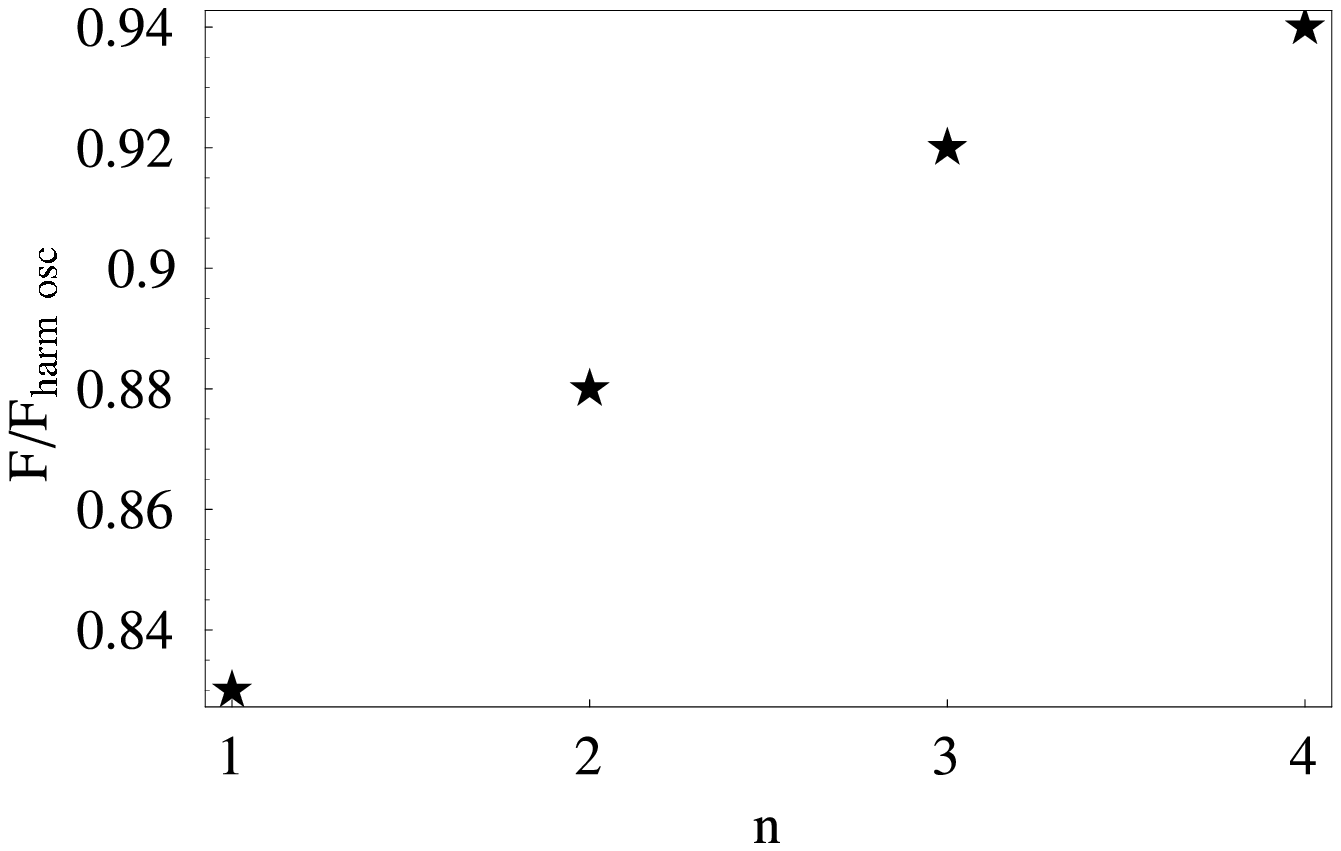}
\caption{\label{fig:ptho}\sl Casimir force for Poschl-Teller
potential in the Dirichlet limit. Exact vs.\ the harmonic
oscillator approximation, eq.~(\ref{eq:ptho}). On the vertical
axis we plotted the ratio of the Casimir force and the harmonic
oscillator approximation $a\to 0$ for $n=1,2,3,4$. For all the
points we chose $m^2$ such that $V(0)+m^2=0$, so $x=0$ is a
turning point.}}

\subsection{$\delta$-function background}

The case of a localized background potential reaches the extreme
limit when $V(x)\to\beta\delta(x)$.  The Casimir force here is a
non-local effect, as can be understood from the fact that the
potential vanishes (almost) everywhere, and the overall sign of
the force cannot be predicted \emph{a priori}.  The WKB
approximation as developed in the previous section includes only
the local semiclassical modification of the Green's function,
which vanishes in this case.  There is a semi-classical method for
this case (\textit{e.g.} as $\beta,\lambda\to \infty$) which sums
over classical paths that reflect from one surface to the
other\cite{SandS, optical1} and accounts successfully for both the
sign and the magnitude of the Casimir force.  By applying the
analysis of Section III.C with $R(k)=\overline R(k)=
\beta/(2ik-\beta)$, we obtain
\begin{equation}
\cF(\lambda,\beta,m,a)= -\frac{\hbar
c}{2\pi}\int_{m}^{\infty}\frac{\kappa
d\kappa}{\sqrt{\kappa^{2}-m^{2}}}\frac{\pp }{\pp a} \ln\left(1
+\frac{\lambda}{2\kappa}\left(1-\frac{\beta}{2\kappa+\beta}e^{-2\kappa
a}\right)\right)
\label{eq40}
\end{equation}
The special case $\beta=\lambda$ has been studied in other
contexts (see eq.~(70) of Ref.~\cite{newapproach}), and our result
agrees for that special case.  For $\lambda\neq \beta$ however no
new physics arises and we will not study that case here. More
interesting is the limit $\lambda,\beta\to\infty$ with $m=0$ in
which case the integral can be performed yielding
\beq
\cF=-\frac{\hbar c\pi}{24 a^2},
\eeq
the correct Casimir force for two impenetrable walls in one
dimension at distance $a$ from each other.

Another interesting limit is $\lambda\to\infty,\, \beta\to 0$ for
$m=0$ which makes contact with the Born approximation of Section
III.D,
\begin{equation}
\left.\cF(\lambda,\beta,m,a)\right|_{m=0,\lambda\to\infty}=
-\frac{\hbar c\beta}{4\pi a}+ \cO(\beta^{2})
\label{eq42}
\end{equation}
which agrees with the result obtained from eq.~(\ref{eq32}). Here
we have a non-local force which is still a buoyancy force.

\section{Beyond one dimension}
\label{sec:ngtr1}

Field theories become more divergent in higher dimensions. Casimir
effects are no exception, indeed they are more problematic because
both the loop divergences and the sharp background divergences
become worse.  Methods for renormalizing the loop divergences, at
least in dimensions where the interaction with the background
field is, in fact, renormalizable, have been worked out in
Refs.~\cite{Graham:2003ib,newapproach,many}.  Divergences arising
from the sharp background can be avoided by smoothing out the
$\delta$-function as described in Ref.~\cite{Graham:2003ib}.

Once the buoyancy has been calculated in one dimension, the
extension to higher dimensions can be constructed by the methods
of Ref.~\cite{Graham:2001dy}.  The core dynamics is the same in
higher dimensions.  The results can be summarized as follows:  All
the results for $n=1$ generalize to any $n<2$ without
complication.  For $2\le n<3$ the generalization succeeds
\textit{except that the limit $\lambda\to\infty$ cannot be taken}.
In that range of $n$ the Casimir buoyancy depends on the cutoff,
$\lambda$, on the strength of the boundary interaction.  Thus
there is  no separation of scales, no ``effective'' low energy
description of Casimir buoyancy.  As $n\to 3$ two further
complications arise:  first, a new counterterm, $\cL_{\rm  CT}=
c_{2}\sigma^{2}$, must be introduced to renormalize the two point
(in the background field, $\sigma$) function.  This means that the
buoyancy depends on a renormalized coupling, actually the mass of
the $\sigma$, that has to be specified.  Second, as suggested by
the appearance of an interaction proportional to $\sigma^{2}(x)$
induced by renormalization, the $\delta$-function gives a
divergent buoyancy.  Instead it is necessary to smooth it out,
replacing it for example, by a Gaussian, as described in detail in
Ref.~\cite{Graham:2003ib}.  The buoyancy force then depends
explicitly on the structure of the surface as well as the shape of
the background.  In the case of a real material one should make
oneself sure of giving a proper description of the reaction of the
material, using for example a plasma model for the metal with
plasma frequency $\omega_{\rm plasma}$. In that case we expect
that, if the $\lambda\to\infty$ and the $\omega_{\rm
plasma}\to\infty$ limits exist then they coincide.

The problem of interest is an $n-1$ dimensional hyperplane
immersed in a medium which is modulated in the
$x_{\perp}$-direction.  As shown in Ref.~\cite{Graham:2001dy}, the
Casimir energy (per unit `area' of the $n-1$ transverse
directions) for such an interface can be written as,
\begin{equation}
\cE^{n}
=\half\int\frac{d^{n-1}p}{(2\pi)^{n-1}}\int_{0}^{\infty}dE\left(
\sqrt{p^{2}+E}-\sqrt{p^{2}+m^{2}}\right)\left(\frac{dN}{dE}-\sum_{k=1}^{M}\frac{dN}{dE}^{(M)}\right)
+ \cE_{\rm  FD}(M)
\label{eq5.11}
\end{equation}
where $\sqrt{E}=\sqrt{k^{2}+m^{2}}$ is the contribution to the
energy from momentum in the $x_{\perp}$ direction.  $dN/dE$ is the
density of states for the one dimensional problem and $
{dN}^{(M)}/{dE}$ is the contribution to the density of states to
$M^{\rm th}$ order in the background field,
$\sigma(x)=\lambda\Delta(x-a)+V(x)$.\footnote{The  ``Levinson''
subtraction discussed at length in Ref.~\cite{Graham:2001dy}, is
unnecessary here since the first Born subtraction does not
contribute.  To carry through the analysis for $n=2$ a Levinson
subtraction would be necessary.}  $\cE_{\rm FD}(M)$ is the
contribution of the Feynman diagrams through $M^{\rm th}$ order in
the background field plus the contributions of counterterms
necessary for renormalization.

The number of Born subtractions is determined by the degree of
divergence of the field theory in $\sigma$ with a $\sigma\phi^{2}$
coupling.  For $n<3$ only the tadpole diagram diverges, so only
the first Born approximation must be subtracted.  For the moment
we set $N=1$ and discuss the extension to $n=3$ at the end of this
section.

The $p$-integration in eq.~(\ref{eq5.11}) is to be understood in
the sense of dimensional regularization, and can be
performed,\footnote{Formally, the integral is defined and
performed for values of ${\rm Re}\ n$ small enough that it
converges.  It is then analytically continued to real, positive
$n$, with careful treatment of singularities encountered along the
way.}
\begin{equation}
\cE^{n}=-\half\frac{\Gamma(-n/2)}{(4\pi)^{n/2}}\int_{0}^{\infty}dE\
E^{n/2} \left(\frac{dN}{dE}- \frac{dN}{dE}^{(1)}\right) + \cE_{\rm
FD}(1)
\label{eq5.2}
\end{equation}
The first Born approximation and the contribution of the tadpole
diagram plus counterterm are linear in $\lambda$ and linear in
$V$.  Therefore the $V$ dependent term they generate is
independent of $a$, so they do not contribute to the buoyancy and
can be dropped,
\begin{equation}
\cF^{n}=
\half\frac{\Gamma(-n/2)}{(4\pi)^{n/2}}\int_{0}^{\infty}dE\ E^{n/2}
 \frac{\p^{2} N}{\p a\p E}
 \label{eq5.3}
\end{equation}
This remarkably simple form is valid for $n<3$, and, of course,
agrees with Section II for $n=1$.

As in one dimension, we take $\sigma(x)=\lambda\delta(x-a)+V$,
construct the Greens function, rotate the integration to the
negative $E$ axis and integrate by parts,
\begin{equation}
\cF^{n}=-\frac{1}{2\pi}\frac{\Gamma(1-\frac{n}{2})\sin{\frac{n\pi}{2}}}
{(4\pi)^{n/2}}\int_{0}^{\infty}dE E^{\frac{n}{2}-1}\frac{\p}{\p
a}\ln\left(1+\lambda G_{0}(a,a,-E)\right).
\label{eq5.4}
\end{equation}
This should be compared with eq.~(\ref{12.22}), which is the
specialization to $n=1$.

Note that the poles in $\Gamma(1-n/2)$ are cancelled by the factor
of $\sin n\pi/2$, which arose from the imaginary part of
$(-E)^{n/2}$.  Thus the only singularities in eq.~(\ref{eq5.4})
arise from divergences at the upper limit of the $E$ integration.
These get more serious as $n$ increases because of the prefactor
of $E^{n/2-1}$.  It is easy to see that the $E$-integral in
eq.~(\ref{eq5.4}) converges for $n<3$:  The high (negative) energy
behavior of $G(a,a,-E)$ is dictated by the WKB approximation,
eq.~(\ref{eq:GWKB}),
\begin{equation}
G(a,a,-E)\to \frac{1}{2\sqrt{E+W(a)}}+\Ord{E^{-1}}
\label{eq5.5}
\end{equation}
(here again $W=m^2+V$) and this behavior leads to a convergent
integral for $n<3$. However, if we attempt to take
$\lambda\to\infty$ before doing the integral, we lose convergence
already at $n=2$.

Beyond eq.~(\ref{eq5.4}) the buoyancy depends on the specific form
of $V(x)$.  To get a feel for its variation, we evaluate the WKB
approximation,
\begin{equation}
\cF^{n}_{\rm WKB}
=\frac{\lambda}{2(4\pi)^{\frac{n}{2}+1}}W^{\frac{n-3}{2}}\frac{dV}{da}\Gamma\left(1-\frac{n}{2}\right)\sin\left(n\frac{\pi}{2}\right)
f\left(\frac{\lambda}{2\sqrt{W(a)}}\right),
\label{eq5.6}
\end{equation}
where
\beq
f(x)=\int_0^\infty dyy^{n/2-1}\frac{1}{y+1}\frac{1}{\sqrt{y+1}+x}.
\eeq
This function can be expressed by means of hypergeometric
functions as
\beq
\label{eq:fhyp}
f(x)=-\frac{\pi}{x}\left(
         {\left( 1 - x^2 \right) }^{\frac{n}{2}-1}
         -1\right) \csc\left(\frac{n\pi }{2}\right)
       + \frac{2}
     {{\sqrt{\pi }}}\Mfunction{\Gamma}\left(\frac{3}{2} - \frac{n}{2}\right)
     \Mfunction{\Gamma}\left(\frac{n}{2}\right)\,
     \Mfunction{F}\left(1;
      \frac{3}{2} - \frac{n}{2},\frac{3}{2};x^2\right)
\eeq
however does not carry much more information than the original
integral, in particular because the interesting limit
$\lambda\to\infty$ translates in $x\to\infty$ where the
hypergeometric function has both a real and an imaginary part. The
imaginary part is cancelled by the first term and the real part is
the one we are interested in but this is difficult to isolate.
However, before carrying on an asymptotic analysis of this
integral it is straightforward to evaluate the $n=2$ result
\beq
\label{eq:f2}
\cF_{\rm
WKB}^{n=2}=\frac{1}{16\pi}\frac{dV}{da}\log\left(1+\frac{\lambda}{2\sqrt{W(a)}}\right)=-\frac{d}{da}\Omega^{n=2}_{\rm
WKB},
\eeq
where the quantum potential is
\beq
\Omega^{n=2}_{\rm
WKB}=-\frac{1}{16\pi}\left(\frac{{\sqrt{W}}\,\lambda }{2} +
  \left( W - \frac{{\lambda }^2}{4} \right) \,
   \log (1 + \frac{2\,{\sqrt{W}}}{\lambda }) -
  W\,\log (\frac{2{\sqrt{W}}}{\lambda })\right).
\eeq
Neither the quantum potential nor the force have a finite limit
when $\lambda\to\infty$ in 2 dimensions. They diverge
logarithmically in $\lambda$.  The case $n=1$ result coincides
with the result of Section II.

Let us now return to the original integral representation to get
an asymptotic expansion for large $x$ (\emph{i.e.\ } large
$\lambda$). Writing it as
\beq
f(x)=\int_0^\infty dz e^{-z x}2\int_1^\infty dy e^{-y
z}\frac{1}{y}(y^2-1)^{n/2-1},
\eeq
performing the integral over $y$, expanding in series for small
$z$ and integrating term by term in $z$ one
gets\footnote{Alternatively one could use the Mellin
representation of the hypergeometric function \cite{Abramowitz}
and move the contour of integration in order to pick the desired
poles. This procedures is much more involved than the one
described here and yields the same results.}
\barr
f(x)&\simeq&
  2\pi \frac{1}{x^{3 - n}}\csc (n\pi ) +
  \frac{1}{x}\pi\csc \left(\frac{n\pi }{2}\right)
     +(2-n)\pi \frac{1}{x^{5 - n}}\csc (n\pi )-\nonumber\\
      &-&
  \frac{\Mfunction{\Gamma}(\frac{1}{2} - \frac{n}{2})
     \Mfunction{\Gamma}(\frac{n}{2})}{{\sqrt{\pi
         }}x^2}+\Ord{x^{-3}}+\Ord{x^{n-7}}.
\label{eq:fasympt}
\earr

This expansion agrees very well with the exact expression
eq.~(\ref{eq:fhyp}) already at $x\simeq 2$, as can be seen in
Figure \ref{fig:ffasympt}, except in the limit $n\to 1$. The
expansion (\ref{eq:fasympt}) has a pole in $n=1$ that goes like
$x^{-4}$. This is not present in the original function $f$ and
indeed adding more terms pushes the pole to higher and higher
powers of $x^{-1}$. However in this case we know the exact result
either from Section II (equation (\ref{eq26})) or by taking the
limit $n\to 1$ of eq.~(\ref{eq:fhyp}) \emph{before} expanding for
$\lambda\to\infty$ (the two limits do not commute).

One thing to notice is that in passing from $n<2$ to $n>2$ the
first term in eq.~(\ref{eq:fasympt}) becomes dominant over the
second one. At exactly $n=2$ they almost cancel each other leaving
behind a term $\log(x)/x$ which reproduces the leading term in
eq.\ eq.~(\ref{eq:f2}) (also the higher order terms agree with the
expansion of eq.~(\ref{eq:f2}) for $\lambda\to\infty$). Another
thing is that from this asymptotic expansion one can see that in
the Dirichlet limit one has a finite force if and only if $n<2$.
For $n<2$ the second term in eq.~(\ref{eq:fasympt}) is dominant
over the first one and hence the force is independent of $\lambda$
\beq
\cF_\infty^{n<2}\simeq\hbar c\frac{dV}{da}\frac{1}{{4(4\pi)}^
    {\frac{n}{2}}}\,W^{\frac{n-2}{2}}\,
    \Mfunction{\Gamma}\left(1 - \frac{n}{2}\right).
\label{eq:fasynls2}
\eeq
Having neglected the first term in eq.~(\ref{eq:fasympt}),
however, this form cannot be used when $n\to 2$. Incidentally
however notice that the limit $n\to 1$ now can be taken safely
(and it gives the usual result eq.~(\ref{eq:WKB0})) since we have
not included the term $\propto \csc(n\pi)/x^{3-n}$ which generated
pole in $n=1$. This pole was a fiction of our procedure of taking
the limit $\lambda\to\infty$ before $n\to 1$. It was signifying
that for $n=1$ the next term in the large $x$ expansion of $f$
decreases slower than $1/x^2$, indeed it is $\propto\log(x)/x^2$.
As we said, including more and more terms in the asymptotic
expansion pushes this pole further and further away.

As anticipated, for $n\geq 2$ the buoyancy force diverges when
$\lambda\to\infty$. Keeping the first term in
eq.~(\ref{eq:fasympt}) one obtains
\beq
\cF^{n>2}_{\infty}\simeq \frac{\hbar c}{4^{n}
    {\pi }^{\frac{n}{2}}}\frac{dV}{da}{\lambda }^{n-2}
    \Mfunction{\Gamma}\left(1 - \frac{n}{2}\right)
    \sec \left(n\frac{\pi}{2}\right).
\label{eq:fasyngt2}
\eeq

Notice that for $n\to 3$ we have a structure $\cF\propto
\frac{1}{n-3}\lambda dV/da$ which comes from a term $\propto
\log{\Lambda}\int d^4x(V(x)+\lambda \delta(x-a))^2$ in the
effective action (here $\Lambda$ is the QFT cutoff). The pole
$1/(n-3)$ from $\sec(n\pi/2)$ is the dimensional regularization
way of seeing the logarithmic divergence $\log\Lambda$ in the
two-legs graphs for $3+1$ dimensions.

\FIGURE{
\includegraphics[width=12cm]{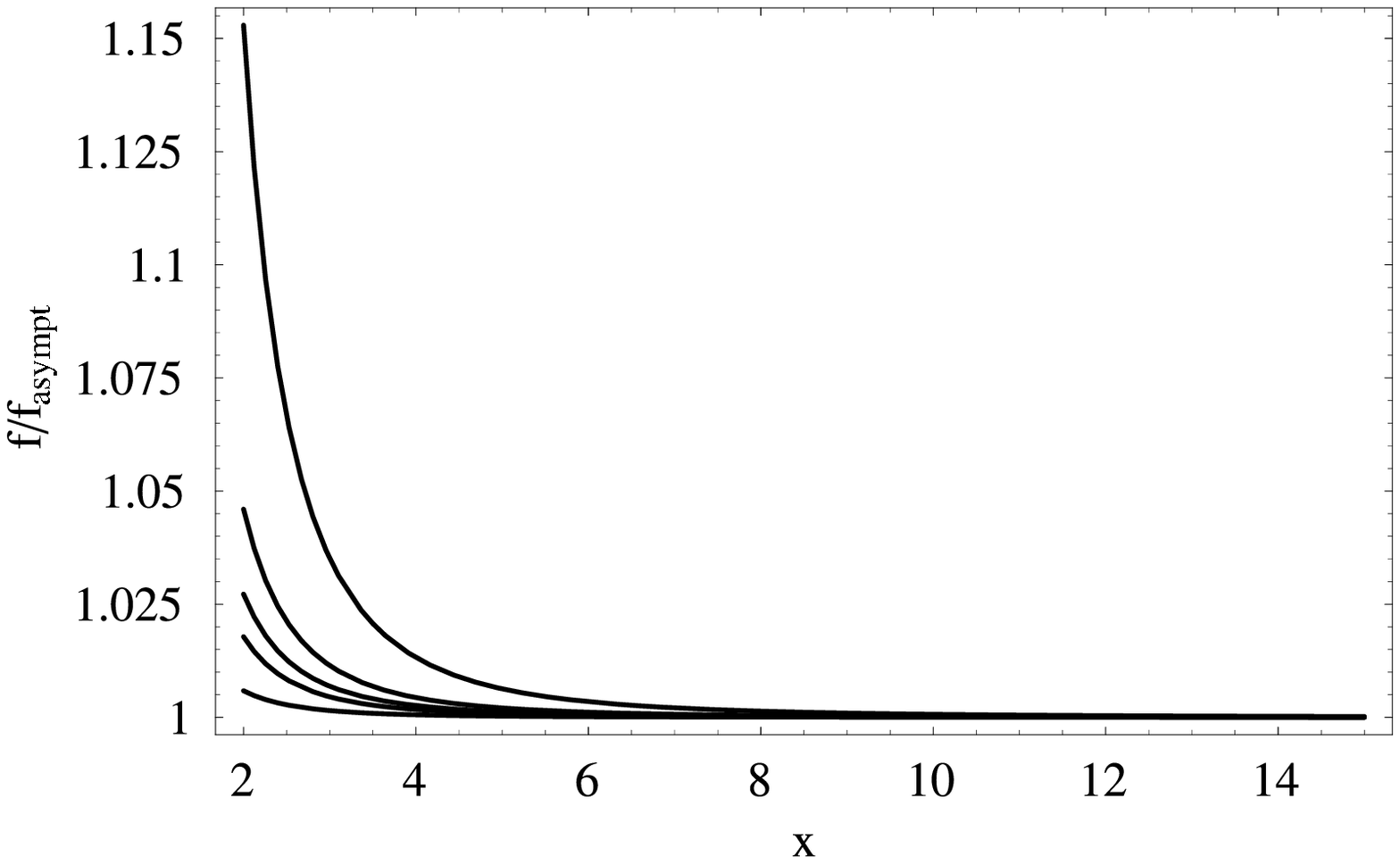}{c}
\caption{\label{fig:ffasympt}\sl Comparison of $f$ in
eq.~(\ref{eq:fhyp}) with its asymptotic expansion
eq.~(\ref{eq:fasympt}) for various $n$ from up to down
$n=1.4,1.8,2,2.2,2.6$.}}

The result eq.~(\ref{eq5.6}) (and its asymptotic expansions
eq.~(\ref{eq:fasynls2}) and eq.~(\ref{eq:fasyngt2})) embodies all
the behaviors we expect from the general analysis. It is a
continuous function of $n$ and finite for $n<3$ and the Dirichlet
limit, $\lambda\to\infty$, can only be taken for $n<2$.

To go all the way to $n=3$, the plane in three dimensional space,
it would be necessary to invoke the full apparatus of the field
theoretic approach to Casimir effects:  the surface must be
smoothed out and the second Born approximation to the density of
states must be subtracted from the $E$-integration and added back
together with the counterterm as a Feynman diagram.

\section{Is Casimir Buoyancy universal?}

In the previous examples we have found Casimir buoyancy is
ubiquitous. However all the previous examples share the fact that
they can be very well described by the zero-order reflection term
in the propagator. The WKB approximation was always an excellent
approximation. There are obvious examples in which this is not
true: Take for example the background $V$ as made of a tall wall
(a tall square wall will do the job) \emph{plus} a shallow
potential, decreasing toward the wall. If the shallow potential
would not be present the delta function would be attracted toward
the wall so for continuity this must be true for sufficiently
small potentials \emph{even if the total potential is decreasing
toward the wall}. This is true because in the expression for the
propagator $\cG_0$ we cannot neglect the influence of the tall
wall (that is why we called it `tall') and this enters only
through a non perturbative term  in the form of the contribution
from a closed path that goes from $a$ to the wall and bounces back
to $a$. More quantitatively by considering the background as made
up of a wall and a smooth $W$, one can expand the propagator as
\cite{Schulman}
\beq
\cG_0(a,a,E)=\frac{i}{2\sqrt{E-W(a)}}+\frac{ie^{-i\pi\mu/2}}{2\sqrt{E-W(a)}}e^{iS_1(a,a,E)}
\eeq
 where $S_1(a,a,E)$ is the action of the closed path that
goes from $a$ to $a$ bouncing on the wall and $\mu$ is the Maslov
index of this orbit ($\mu=1$ is the wall is `smooth' on
frequencies $\sqrt{E}$ and $\mu=2$ for otherwise `hard' walls).
Inserting into eq.\ (\ref{eq:forceE}) one finds a general formula
including both local (e.g.\ $W$ and its derivatives) and non-local
(depending on integrals of $W$ or independent of $W$)
contributions. The two effects interplay and they are difficult to
separate but it is easy to extract the two limits: 1) When $W\to
0$ one recovers the Casimir attraction between impenetrable plates
(compare with eq.\ (\ref{eq40}) with $\beta\to\infty$) 2) When the
wall is far away ($S_1\gg 1$) the second term is exponentially
smaller than the first and we find buoyancy from the first term
only.

In general then a more correct statement would be that
\emph{Casimir buoyancy will occur in the cases where non-local
effects, like closed orbits contributions to the propagator, are
negligible.} We believe we have presented enough examples here to
convince the reader that this is not a too stringent request.

\section{Acknowledgements}
A.~S.\ is a Bruno Rossi Fellow (partially supported by INFN) and
Jonathan A.~Whitney Fellow.  R.~L.~J.\ thanks P.~Hoodbhoy for some
early conversations on this subject.  This work is also supported
in part by funds provided by the U.S.~Department of Energy
(D.O.E.) under cooperative research agreement DE-FC02-94ER40818.

We believe that the response of a single boundary to an
inhomogeneous background field was first considered for the case
$V(x)\propto |x|$ in a paper begun in collaboration with one of
the present authors (RLJ) Ref.\cite{PH}.

\bibliographystyle{JHEP}

\begin{thebibliography}{99}
\bibitem{casimir}
H.~B.~G.~Casimir,
Kon.\ Ned.\ Akad.\ Wetensch.\ Proc.\  {\bf 51}, 793 (1948).
\bibitem{Jaekel}
  M.~T.~Jaekel and S.~Reynaud,
  Phys.\ Lett.\ A {\bf 167}, 227 (1992)
  [arXiv:quant-ph/0101080].
\bibitem{BirrelDavis} N.~D.~Birrell and P.~C.~W.~Davies, \emph{Quantum fields in curved
space}, Cambridge Univ.~Press, 1982.
\bibitem{dirichlet}  Both kinds of divergences are discussed in detail in Ref.~\cite{Graham:2003ib}
\bibitem{eft} See for example S.~Weinberg, \emph{The Quantum Theory of Fields},
Cambridge Univ.~ Press, 1996.
\bibitem{Graham:2003ib}
N.~Graham, R.~L.~Jaffe, V.~Khemani, M.~Quandt, O.~Schroeder and
H.~Weigel,
Nucl.\ Phys.\ B {\bf 677}, 379 (2004) [arXiv:hep-th/0309130].
\bibitem{Graham:2001dy}
N.~Graham, R.~L.~Jaffe, M.~Quandt and H.~Weigel,
Phys.\ Rev.\ Lett.\  {\bf 87}, 131601 (2001)
[arXiv:hep-th/0103010]; Annals Phys.\  {\bf 293}, 240 (2001)
[arXiv:quant-ph/0104136].
\bibitem{Zorbas} J.~Zorbas, Journ.\ Math.\ Phys. {\bf 21}, 840
(1980).
\bibitem{texts} R.~Newton, \emph{Scattering Theory of Waves and Particles} McGraw-Hill, New York, 1966; M.~Goldberger and K.~Watson,
\emph{Collision Theory} John Wiley \& Sons, New York, 1964;
K.~Gottfried \emph{Quantum Mechanics} W.~A.~Benjamin, New York,
1966.
\bibitem{landau}
L.~D.~Landau and E.~M.~Lifshitz, \emph{Quantum Mechanics:
Nonrelativistic Theory} Pergamon Press, Oxford, England, 1997.
\bibitem{BerryMount} M.~V.~Berry and K.~E.~Mount, Reps.~Prog.~Phys.\ {\bf 35}, 315 (1972).
\bibitem{Abramowitz} M.~Abramowitz and I.~Stegun, \emph{Handbook of Mathematical Functions}, Dover Publications, Inc., New York, 1970.
\bibitem{optical2}
A.~Scardicchio and R.~L.~Jaffe, \emph{The optical approach to
Casimir energy: Local observables and thermal corrections}, to
appear.
\bibitem{classicaltemp} M.~Bordag, U.~Mohideen
and V.~M.~Mostepanenko, Phys.~Rept.~{\bf 353}, 1 (2001)
    [arXiv:quant-ph/0106045].
\bibitem{optical1}
R.~L.~Jaffe and A.~Scardicchio, Phys.\ Rev.\ Lett.\  {\bf 92},
070402 (2004) [arXiv:quant-ph/0310194]; A.~Scardicchio and
R.~L.~Jaffe,
Nucl.\ Phys.\ B {\bf 704}, 552 (2005) [arXiv:quant-ph/0406041].
\bibitem{Hilberttransf}
R.~Bracewell, \emph{The Hilbert Transform} in \emph{The Fourier
Transform and Its Applications}, 3rd ed.\ New York. McGraw-Hill,
pp.\ 267-272, 1999.
\bibitem{many}
E.~Farhi, N.~Graham, P.~Haagensen and R.~L.~Jaffe,
Phys.\ Lett.\ B {\bf 427}, 334 (1998) [arXiv:hep-th/9802015];
N.~Graham, R.~L.~Jaffe, V.~Khemani, M.~Quandt, M.~Scandurra and
H.~Weigel,
Phys.\ Lett.\ B {\bf 572}, 196 (2003) [arXiv:hep-th/0207205]. For
a review, see N.~Graham, R.~L.~Jaffe and H.~Weigel,
Int.\ J.\ Mod.\ Phys.\ A {\bf 17}, 846 (2002)
[arXiv:hep-th/0201148].
\bibitem{SandS} M.~Schaden and L.~Spruch, Phys.~Rev.~A {\bf 58}, 935
  (1998); Phys.~Rev.~Lett. {\bf 84}, 459 (2000).
\bibitem{newapproach}
N.~Graham, R.~L.~Jaffe, V.~Khemani, M.~Quandt, M.~Scandurra and
H.~Weigel,
Nucl.\ Phys.\ B {\bf 645}, 49 (2002) [arXiv:hep-th/0207120].
\bibitem{Schulman} L.~S.~Schulman, \emph{Techniques and Applications of Path Integration},
Wiley--Interscience, New York, 1981.
\bibitem{PH}
P.~Hoodbhoy,
[arXiv:quant-ph/0411031].


\end{thebibliography}

\end{document}